\documentclass[nofootinbib,a4paper,aps,prd,10pt,superscriptaddress,showpacs,twocolumn]{revtex4}

\usepackage{braket}
\usepackage{float}
\usepackage{graphicx}
\usepackage{csquotes}
\usepackage{multirow}
\usepackage{natbib}
\usepackage{pifont}
\usepackage{amssymb}
\usepackage{bm}
\usepackage{amsfonts}
\usepackage{amsmath}
\usepackage{amssymb}
\usepackage[titletoc]{appendix}
\usepackage{color}
\usepackage{hyperref}
\usepackage{cleveref}
\usepackage[english]{babel}
\usepackage{booktabs}
\usepackage{mathtools}
\usepackage{subfigure}
\usepackage{enumerate}
\usepackage[utf8]{inputenc}
\usepackage{newunicodechar}
\newunicodechar{−}{\textminus}

\definecolor{vividviolet}{rgb}{0.62, 0.0, 1.0}
\definecolor{amaranth}{rgb}{0.9, 0.17, 0.31}
\definecolor{palatinateblue}{rgb}{0.15, 0.23, 0.89}
\definecolor{brightpink}{rgb}{1.0, 0.0, 0.5}
\definecolor{cornflowerblue}{rgb}{0.39, 0.58, 0.93}
\definecolor{deepcarminepink}{rgb}{0.94, 0.19, 0.22}
\definecolor{radicalred}{rgb}{1.0, 0.21, 0.37}

\hypersetup{ linktoc=all,
    colorlinks, linkcolor={palatinateblue},
    citecolor={brightpink}, urlcolor={amaranth}
}

\def\be{\begin{equation}}
\def\ee{\end{equation}}

\begin{document}
\title{Bondi accretion disk luminosity around neutral and charged Simpson-Visser spacetimes}
\author{Serena Gambino}
\email{s.gambino@ssmeridionale.it}
\affiliation{Scuola Superiore Meridionale (SSM), Largo San Marcellino 10, 80138 Napoli, Italy.}
\affiliation{Istituto Nazionale di Fisica Nucleare (INFN), Sezione di Napoli, Napoli, Italy.}
\author{Roberto Giambò}
\email{roberto.giambo@unicam.it}
\affiliation{Universit\`a di Camerino, Via Madonna delle Carceri 9, 62032 Camerino, Italy.}
\affiliation{INFN, Sezione di Perugia, Perugia, 06123, Italy.}
\affiliation{INAF - Osservatorio Astronomico di Brera, Milano, Italy.}
\author{Orlando Luongo}
\email{orlando.luongo@unicam.it}
\affiliation{Universit\`a di Camerino, Via Madonna delle Carceri 9, 62032 Camerino, Italy.}
\affiliation{Department of Nanoscale Science and Engineering, University at Albany-SUNY, Albany, New York 12222, USA.}
\affiliation{Istituto Nazionale di Fisica Nucleare (INFN), Sezione di Perugia, Perugia, 06123, Italy.}
\affiliation{INAF - Osservatorio Astronomico di Brera, Milano, Italy,}
\affiliation{Al-Farabi Kazakh National University, Al-Farabi av. 71, 050040 Almaty, Kazakhstan.}
\begin{abstract}
We investigate relativistic Bondi accretion in the Simpson-Visser spacetime, which, via a single parameter $\ell$, interpolates between the Schwarzschild, regular black hole, extremal and wormhole regimes. First, we analyze the neutral Simpson-Visser geometry, recovering Schwarzschild at $\ell=0$,  and then its charged extension of the Reissner-Nordstr\"om metric. In both these cases, we derive the conservation equations and analyze two representative fluid models: a barotropic perfect fluid and a constituent with an exponential density profile. By varying the parameters across regimes, we locate critical (sonic) points and integrate velocity, density and pressure profiles. While near-horizon inflow velocities are similar across the different solutions, we find that the critical radius and the resulting accretion rates and luminosities severely change, depending on the value of the parameter and  type of fluid. Remarkably, the barotropic and exponential cases exhibit different trends in the outer regions. Moreover, by extending the analysis to the charged SV spacetime, we find that the presence of a central charge $Q$ produces additional, albeit modest, shifts in the sonic radius which, in combination with those induced by the regularization parameter $\ell$, could provide a double observational marker. In particular, while $\ell$ acts predominantly on the position of the critical point, in the barotropic fluid case, the electromagnetic contribution of $Q$ slightly dampens the inflow velocity near the horizon.
\end{abstract}


\maketitle

\tableofcontents
%
%
%
\section{Introduction}
\label{INTRODUCTION}

Black holes (BHs) are among the most intriguing objects predicted by general relativity and solutions to Einstein's equations. While many observations in recent years have confirmed their existence, their nature remains unclear, as highlighted by recent gravitational wave detections \cite{LIGO} and the first BH shadow image \cite{EHT}. Nevertheless, the classical singularities predicted at their cores remain a significant theoretical challenge.

Indeed, the classical theorems of Penrose and Hawking \cite{Hawking:1970,Hawking:1996}, under rather general assumptions, predict that spacetime is geodesically incomplete - a condition that, in most cases, implies the presence of spacetime singularities.

Several hypotheses concerning BH solutions that do not predict the central singularity have been proposed over the years. These are called regular BHs (RBHs) and were initially proposed mathematically by Sakharov and Gliner \cite{Sakharov:1996,Gliner:1996}. Bardeen was the first to attempt to construct a RBH \cite{Bardeen:1968}, using the hypothesis of a topological charge to eliminate the singularity. Many others approaches have followed after this pioneering results, see e.g. \cite{Lan:2023cvz,Bueno:2024dgm,Giambo:2023,Corona:2024}.

Alternative models have been developed to address this issue. Indeed, RBHs evade central singularities while maintaining event horizons, thereby complying with Penrose's cosmic censorship hypothesis in a modified context (see, for example \cite{Hayward:2005,Fan-Wang:2016,Dymnikova:2004,Ansoldi:2008jw,Bronnikov:2001,Fan:2016hvf,Kurmanov:2024}). These geometries offer a physically motivated resolution to singularities and belong to a broader class of BH mimickers \cite{Olmo:2023,Rosa:2022,Mazza:2021rgq}, comprising objects that resemble BHs externally but differ internally (see also \cite{Shaikh:2019hbm,Bambi:2025wjx,Carballo-Rubio:2025fnc,Bambi:2023try}).

In this work, we focus on a particular family of RBHs: the Simpson-Visser (SV) solutions \cite{Simpson:2018}. This one-parameter, spherically symmetric solution continuously interpolates between four regimes, depending on the regularization parameter $\ell$: RBHs, black bounce solutions and traversable wormholes plus, when these parameter vanishes, the standard Schwarzschild BH is recovered. Particularly, previous investigations have examined Novikov-Thorne thin accretion disks in the context of the SV metric, revealing subtle variations in luminosity and temperature profiles compared to Schwarzschild BHs \cite{Bambhaniya:2021,Murodov:2024oym}.

Further investigations have addressed various physical aspects of the SV spacetime, including models with nonlinear electrodynamics \cite{Alencar:2025,Balart:2014jia,Kruglov:2016ezw,Kruglov:2017fck,Gullu:2020ant}, general relativistic magnetohydrodynamics  simulations \cite{Guerrero:2022,Combi:2024ehi}, particle dynamics \cite{Silva:2024fpn}, linear perturbations \cite{Franzin:2023slm,Lobo:2020kxn}, and implications for BH observables \cite{Duran-Cabaces:2025sly,Lu:2024nev,Yang:2024mro,Luongo:2025iqq}. Hence, here we focus on investigating the Bondi accretion in the SV metric \cite{Bondi:1952,Michel:1972} and investigate how two distinct matter energy-momentum tensors produce observationally distinguishable features when accreted onto RBH geometries compared to standard BHs. More precisely, we employ two main fluids,
\begin{enumerate}
    \item[-] a pure barotropic fluid, whose equation of state, $w$, does not depend on density, $\rho$, and entropy, namely $P = w\,\rho$, with $P$ the pressure. Moreover, we ensure $w$ to be a  constant,  representing a wide class of accreting matter;
    \item[-] a fluid with an exponential density profile \cite{Sofue:2013a,Sofue:2013b}, which depends solely on the radial coordinate. This model has previously been used to model dark matter environments in accretion scenarios \cite{Boshkayev:2020,Sofue:2008wt,Sofue:2011kw}.
\end{enumerate}
We apply both to a selection of solutions embedded within the SV framework to calculate key observables, such as the accretion rate, critical radius and resulting luminosity. Afterwards, we explore the impact of a charge $Q$, investigating the charged SV metric that can be compared with the Reisser-Nordstr$\ddot o$m (RN) spacetime, extending it through the parameter $\ell$, enabling regularity \cite{Franzin:2023slm}, thereby obtaining a deformation of electrically charged BHs and wormholes, studying how $Q$ modifies the location of the sonic point, the mass-flux and luminosity profiles for both barotropic and exponential-density fluids. Furthermore, our results are compared with Schwarzschild and RN metrics and we thus show that, while the Bondi formalism is stable under exponential profiles, the barotropic case exhibits shifts in the critical point, as well as a different trends in the mass flux and luminosity, compared to the exponential case. Observational consequences of our recipe are therefore discussed and commented in view of future observations.

The paper is organized as follows. In Sect. \ref{BONDISV}, we set up the framework for uncharged SV spherical accretion and, moreover, the class of BH solutions, under exam. In addition, we present the general Bondi formalism for spherically symmetric accretion in the SV spacetime. Afterwards, Sect. \ref{FLUIDS} defines the two fluid models and their key relations, adopted throughout the text. In Sect. \ref{RN_GENERAL_FORMALISM}, we derive the same relations but for the case of charged SV metric. Sects. \ref{RESULTS} and \ref{RN_RESULTS} are devoted to our numerical findings for uncharged and charged SV solutions, respectively. Finally, in Sect. \ref{CONCLUSION} we discuss the implications and future directions of this study\footnote{Throughout the text, we use geometric units, namely $G = c = \hbar = 1$.}.

%
%
\section{Sperically symmetric accretion in the Simpson-Visser spacetime}
\label{BONDISV}

We will now illustrate how we model the accretion of two fluids in a spherically symmetric, relativistic setting. First, we introduce the SV solution, which for the moment does not include an electric charge, i.e. \emph{neutral SV spacetime}. Then, using the standard procedure, we derive the accretion equation for these specific geometries, leaving the parameter $\ell$ unspecified. We also examine how the critical point changes through the two analysed fluids.

\subsection{Theoretical set up}
\label{SV_METRIC}

The SV metric \cite{Simpson:2018} is a one-parameter regularization of the Schwarzschild solution, introduced in terms of an auxiliary physical coordinate $x$, whose relationship with the usual spherical radial coordinate $r$, is given by

\begin{equation}
\label{SV_coordinate}    r(x)\equiv\sqrt{x^2 + \ell^2}\,,
\end{equation}

and a constant parameter $\ell \geq 0$.

Accordingly, the line element acquires the following form,

\begin{align}
\label{SV_general_metric}
  ds^2 &= -A(x)dt^2 + \frac{1}{B(x)}dx^2 + C(x) d\Omega^2 \nonumber \\
       &= -\left(1-\tfrac{2M}{\sqrt{x^2 + \ell^2}}\right)\,dt^2 + \frac{dx^2}{1-\tfrac{2M}{\sqrt{x^2 + \ell^2}}} + (x^2 + \ell^2)\,d\Omega^2\,,
\end{align}
where $d\Omega^2\equiv d\theta^2+\sin^2\theta d\phi^2$. This particular solution defines four distinct cases, determined by the parameter $\ell$, namely:

\begin{itemize}
  \item[-] $\ell\rightarrow0$: the Schwarzschild solution recovered with $A=B=1-2M/r$ for $r>2M$,
   \item[-] $0<\ell<2M$: a RBH with two horizons at $r=\sqrt{4M^2-\ell^2}$,
   \item[-] $\ell=2M$: an extremal BH with a double horizon at $r=2M$,
   \item[-] $\ell>2M$: a traversable wormhole throat located at $r=\ell$ (or $x=0$).
\end{itemize}

\subsection{Bondi relativistic accretion}
\label{SV_GENERAL_FORMALISM}

Here, we use the general form of the SV metric, as defined in Eq. \eqref{SV_general_metric}, for a generic static spherically-symmetric spacetime, to derive the key equations for modeling relativistic Bondi accretion in two fluids.

First, we need the definition of the energy-momentum tensor to define the accretion of the fluids around the BH. For this purpose, we use the definition of a perfect fluid, whereby the energy-momentum tensor becomes
\begin{equation}
\label{stress_energy_tensor}
T_{\mu\nu} = (P + \rho) u_{\mu} u_{\nu} + P g_{\mu\nu},
\end{equation}
where $\rho$ is the energy density, $P$ the pressure and $ u^\mu = u^\mu(x) $ is the four-velocity vector. Thanks to the symmetries of the problem, the four-velocity reduces to the following non-zero components:
\begin{equation}
\label{four_velocity_components}
u^\theta = 0 \,\,\, \text{and} \,\,\, u^\phi = 0  \implies u^{\mu} = \frac{dx^{\mu}}{d\tau} = (u^t, u^x, 0, 0)\,,
\end{equation}
where $ \tau $ is the proper time. In this type of study, usually the time coordinate $ u^t $ tends to be expressed in terms of the radial coordinate and this time we express it in terms of the coordinate $x$ introduced before, i.e. in terms of $ u^x $. The expression is, therefore, the one that bonds the two components,
\begin{equation}
\label{normalization_condition} u_{\mu} u^{\mu} = -A(x) (u^t)^2 +\frac{1}{B(x)} (u^x)^2 = -1,
\end{equation}
and, from it, we immediately obtain\footnote{For the sake of simplicity, notice that we also rename the coordinate by $u^x=u$.}
\begin{equation}
\label{time_component_velocity} u^t = \pm \sqrt{\frac{u^2 + B(x)}{A(x) B(x)}}\,.
\end{equation}
Above, we select the solution with a positive sign, ensuring that $u^t>0$ and guaranteeing causality, at the same time.

Focusing on the $u$ component, which is related to the radial coordinate $r$ by the equation in Eq. \eqref{SV_coordinate} allows us to describe two different physical processes, depending on the sign of $u$. More precisely, an outgoing flow is represented by $u>0$, while an inflow, characteristic of accretion, is represented by $u<0$ \cite{Bahamonde:2015,Debnath:2015}.
To characterize relativistic fluid accretion, we select the equatorial plane, i.e., $ \theta = \pi/2 $, leading to the metric determinant, $\sqrt{-g} = x^2 + \ell^2$.

At this stage, we therefore require to investigate the conservation equations and critical points of our problem, as below reported.

\subsubsection*{Conservation equations}

\begin{itemize}

\item[-] The first relation used to model Bondi relativistic accretion is the conservation of the flux of energy density. This condition is realized by the following relation

\begin{equation}
T^{\mu}{}_{t;\mu} =\frac{1}{\sqrt{-g}}\, \partial_x\!\left(\sqrt{-g}\,T^{x}_{\ \ \ t}\right)=0\,,
\end{equation}

in which we already specify the component of the stress-energy tensor we are going to use, i.e. $\nu = t$.

Upon integration and by lowering an index\footnote{To clarify: $\frac{1}{\sqrt{-g}}\, \partial_x\!\left(\sqrt{-g}\,T^x_{\,\, t}\right)=0 \rightarrow \left(\sqrt{-g}\,T^x_{\,\, t}\right)= \mathcal{C}_1$.}, this quantity yields the continuity equation
\begin{equation}
\label{1}
(P + \rho)  u \sqrt{u^2 + A(x)} (x^2 + \ell^2) = \mathcal{C}_1\,,
\end{equation}
where $\mathcal{C}_1$ is an integration constant.

Please note that, in this relation and in the above and following ones, $u^t$ is irrelevant while $u$ becomes the quantity to be dealt with later. Additionally, as already did for Eq. \eqref{1} we will exploit in all the derivation the fact that $A(x)=B(x)$.

\item[-] The second conservation equation we consider, is the mass flux conservation $J^{\mu}$ is given by:
\begin{equation}
\label{mass_flux_conservation}
J^{\mu}\! _{;\mu} = (\rho u^\mu)_{;\mu} = \frac{1}{\sqrt{-g}} \partial_x \left( \sqrt{-g} J^x \right) = 0\,,
\end{equation}
and, after the integration, we have
\begin{equation}
\label{3}
\rho u (x^2+\ell^2)  = \mathcal{C}_3\,,
\end{equation}
with $ \mathcal{C}_3 $ another constant.
\end{itemize}

Additionally, combining Eqs. \eqref{1} and \eqref{3}, we derive the Bernoulli equation:
\begin{equation}
\label{5}
\frac{P + \rho}{\rho} \sqrt{u^2 + A(x)} = \mathcal{C}_4,
\end{equation}
with $\mathcal{C}_4$ a constant determined by $\mathcal{C}_1$ and $\mathcal{C}_3$. The last relation will be useful in the following section, as it will be used to derive \emph{the equation for critical points}.

\subsubsection*{Critical points}
\label{critical_point_analysis}

Within spherically symmetric accretion and,  in particular, within Bondi accretion, a critical point, $r_c$, can be identified at which the fluid transitions from a subsonic to a supersonic regime. This transition occurs when the radial velocity equals the speed of sound, $u^2 = c_s^2$, a phenomenon known as \emph{sonic point}.

Indeed, in the non-relativistic version of the accretion process, these two points are often synonymous. However, in the relativistic extension of the Bondi accretion, the fluid can become supersonic prior to reaching this critical point. Hence, the two points do not always coincide, see e.g.  \cite{FreitasPacheco:2011,Richards:2021zbr,John:2019rhj}). However, it provides information on how the fluid behaves in the accretion disk and how its properties change.

In this work, we denote the critical point $x_c$, for consistency in notation.

Accordingly, we perform a critical point analysis following Ref. \cite{Michel:1972}. To do so, we require a combination of all the other conservation laws, as was done in previous works, see Refs.  \cite{Debnath:2015,Babichev:2004}. First, we define a new, useful, variable as
\begin{equation}
\label{variable_V}  V^2 = \frac{d \ln(P + \rho)}{d \ln \rho} - 1\,.
\end{equation}
which mimics a sound speed relation.

Afterwards, we take the logarithmic derivatives of Eqs. \eqref{3}-\eqref{5}, combining them and using the variable $V^2$ in Eq. \eqref{variable_V}.

By further simplifications, we thus obtain\footnote{Here, we did not use the explicit expression of the metric function $A(x)$ just to simplify the overall notation.}
\begin{align}
\label{critical_point_eq}
   & \left[ V^2 - \frac{u^2}{u^2 + A} \right] \frac{du}{u} +
   \left[ \frac{2x}{(x^2+\ell^2)} V^2 - \frac{A'}{2\left(u^2 + A\right)} \right] dx = 0\,,
\end{align}
recalling that $A^\prime(x)=\frac{2Mx}{(x^2+\ell^2)^{3/2}}$.

The critical point is then determined when both of the bracketed terms go to zero simultaneously, leading to the following system of equations
\begin{subequations}
\begin{align}
\label{crit_point_conditions}
    & V_c^2 = \frac{u_c^2}{u_c^2 + A(x_c)} \,, \\
    & \frac{2x_c}{(x_c^2+\ell^2)} V_c^2 - \frac{A'(x_c)}{u_c^2 + A(x_c)} = 0 \,.
\end{align}
\end{subequations}
After some calculations, we then obtain the following set of expressions for the two quantities $ u_c^2 $ and $ V_c^2 $
\begin{subequations}

\begin{align}
&\, \label{vc2_expression} V_c^2 =\frac{1}{\frac{2\sqrt{x^2+\ell^2}}{M} - 3}\,, \\
&\, \label{uc2_expression} u_c^2 = \frac{M}{2\sqrt{x^2+\ell^2}} \,.
\end{align}
\end{subequations}

These equations, along with the expressions of their general counterparts evaluated at the critical point,  enable us to study the critical point, $x_c$, and consequently to constrain the integration constants.

%
%
\section{The accreting fluids}
\label{FLUIDS}

We select two accreting fluids, among the most relevant ones. Precisely, we first adopt barotropic fluids, i.e. governed by the equation of state (EoS)
\begin{equation}
\label{equation of state}
P(x) = w \rho(x),
\end{equation}
where $w=const$ is the state parameter or barotropic factor.

Second, we explore an exponential, and quite practical, density shape, also known as \emph{Sofue profile} \cite{Sofue:2013a,Sofue:2013b}:

\begin{equation}
\label{Sofue_density}
    \rho(r) = \rho_0 e^{-\sqrt{x^2 + \ell^2}/r_0} \,,
\end{equation}
with $\rho_0$ is a characteristic density and $r_0$ is the core radius and we already shown the expression with the change of coordinate described in Eq. \eqref{SV_coordinate}.

On the one hand, the first choice deals with the perfect analogy with cosmology, where barotropic fluids appear quite ubiquitous, i.e., spanning from background cosmology \cite{Alfano:2025gie,Alfano:2024fzv,Alfano:2024jqn,Carloni:2024zpl,Luongo:2024fww}, up to early times \cite{Anaya-Galeana:2024wst} and passing through contexts of Hubble tension \cite{Carloni:2025jlk} and unified dark energy models \cite{Dunsby:2024ntf,Dunsby:2023qpb,Luongo:2014nld,Dunsby:2016lkw}. On the other hand, the second choice is motivated by recent literature that has raised direct evidences in favor of such a profile, see e.g. \cite{Boshkayev:2020,Kurmanov:2021,Boshkayev:2022vlv,Boshkayev:2021chc}, plus evident advantage to handle dark matter in spirals \cite{2015PASJ...67...75S}.

At this stage, we thus explore both the scenarios, emphasizing the corresponding results

\subsection{Bondi accretion for a barotropic fluid}
\label{variables_barotropic}

Let us start by studying the case of the barotropic fluid. Using the integral relations derived in Sect.  \ref{SV_GENERAL_FORMALISM}, we can express the unknown variables in terms of metric functions and integration constants. This method, outlined in Ref. \cite{Debnath:2015,Bahamonde:2015}, allows us to express the mass accretion rate $\dot{M}$ in terms of these variables, as previously discussed in Ref. \cite{Babichev:2004}.

Since we have an EoS, Eq. \eqref{equation of state}, we have a relation between pressure and density, and therefore we can easily substitute $P$. By virtue of this, we are able to isolate $u$ from Eq. \eqref{1} and to obtain the velocity
\begin{equation}
\label{velocity_expression_barotropic}
u = \pm \sqrt{\frac{\mathcal{C}_4^2}{(1+w)^2}-A(x)} \,.
\end{equation}
Here, we choose the negative solution since, as we previously said, we need to simulate an inflow of matter.

The last variable we need to close the system is the energy density $\rho$. Starting by Eq. \eqref{5} and substituting the velocity from Eq. \eqref{velocity_expression_barotropic} we arrive to the following expression
\begin{equation}
\label{density_expression_barotropic}
\rho = \frac{(1+w)\mathcal{C}_1}{\mathcal{C}_4\sqrt{\mathcal{C}_4^2 - A(x)(1+w)^2}}\,.
\end{equation}
Both Eqs. \eqref{velocity_expression_barotropic} and \eqref{density_expression_barotropic} depend on some integration constants, that can be singled out as we will show in Sect. \ref{RESULTS}. From their magnitudes, it is thus possible to fix reasonable bounds over the observable quantities that we are going to constrain.

\subsubsection*{Critical points}

To find the critical point, we recall what we previously discuss and equate the expression of the velocity at the critical point, Eq. \eqref{uc2_expression}, with its generic counterpart for the barotropic case, Eq. \eqref{velocity_expression_barotropic}, evaluated at $x=x_c$, and since these two quantities must be equal, we can extract the expression for $x_{\rm c}$. The result reads
\begin{equation}
\label{crit_point_barotropic} x_{\rm c} = \pm \sqrt{ \left[\frac{3M^2 (1+w)^2}{2(\mathcal{C}_4^2 - (1+w)^2)}\right] - \ell^2 }\,,
\end{equation}
which depends on both $\ell$ and $w$, i.e., on the type of matter chosen to accrete.

\subsection{Bondi accretion with exponential density profile}
\label{Variables_sofue}

Following the previous  derivation, computed in the barotropic case, we now work the exponential density profile out, from  \eqref{Sofue_density} and start deriving the velocity and pressure profiles.

This case differs physically from the barotropic fluid since the density profile is already known and, therefore, we do not need to evaluate it.

More precisely, starting from Eq. \eqref{3}, we express the velocity by explicating $u$ in the following form

\begin{equation}
\label{exponential_velocity_expression} u = \frac{\mathcal{C}_3}{\rho (x^2+\ell^2)}\,.
\end{equation}
We can now plug the above  expression  into Eq. \eqref{5} to determine the pressure profile, yielding
\begin{equation}
\label{exponential_pressure_expression} P = \left[ \frac{\mathcal{C}_4}{ \sqrt{\left(\frac{\mathcal{C}_3^2}{\rho^2(x^2+\ell^2)} + A(x) \right)}}-1\right] \rho \,.
\end{equation}
%


\subsubsection*{Critical points}

We can now find the expression for critical points, always adopting the same procedure as in the case of barotropic fluids.

From the equality between the two velocity expressions, we have
\begin{equation}
  \label{crit_point_exp} x_{\rm c} = \pm \sqrt{ \left(\frac{2\mathcal{C}_3^2}{M \rho_{\rm c}^2}\right)^{2/3}  - \ell^2 }\,.
\end{equation}
In the above relation, we have a dependence of the radius on the right side of the expression in the term $\rho_{\rm c}$. Accordingly, this expression cannot be solved analytically to find the critical radius. Hence, we perform numerically, as it will be explained later in Sect.  \ref{RESULTS}.

\subsection{Mass accretion rate and disk luminosity}
\label{SEC:BONDI_exp_accretion_rate}

Let us now discuss the formula for Bondi accretion, starting from the mass accretion rate $\dot{M}$, that is derived from the conservation of mass flux. We integrate the mass flux over a two-dimensional surface \cite{Debnath:2015} and obtain
\begin{equation}
\label{accretion_rate}
\dot{M} = - \int d\theta d\phi \sqrt{-g} T^x\! _0\,,
\end{equation}
that can be recast as
\begin{align}
    \label{accretion}
    \dot{M} &= -4\pi (x^2 + \ell^2) (P(x) +\rho(x)) u\sqrt{u^2+A(x)}\,.
\end{align}

To find the correct expression to describe the luminosity, we employ the Eddington luminosity, i.e., representing a limit that indicates the maximum luminosity above which radiation pressure exceeds gravitational attraction, thereby halting the accretion process \cite{Frank:2002}.

The Eddington luminosity is derived by equating the gravitational force acting on a proton with the radiation pressure force exerted by a luminous source. As radiation primarily interacts with electrons via Thomson scattering, electrostatic coupling ensures that the entire plasma feels the effect.

Hence, we consider
\begin{equation}
\label{Eddington_lum}
L_{\rm Edd} = \frac{4\pi GM m_p}{\sigma_T},
\end{equation}
where $ m_p $ is the proton mass and $ \sigma_T $ is the Thomson cross-section.

To account for the efficiency through which the accreted mass is converted into radiation, we introduce an efficiency parameter, labeled $ \eta_{\rm eff} $. The final expression for the luminosity turns out to be  \cite{Pfahl:2000si}
\begin{equation}
\label{bondi_luminosity}
L = \eta_{\rm eff} \dot{M}\,,
\end{equation}
with $\dot{M}$ given by Eq. \eqref{accretion}.

%
%
\section{Sperically symmetric accretion in charged Simpson-Visser geometry}
\label{RN_GENERAL_FORMALISM}

Genealizing the SV spacetime to include electric charge appears the simplest attempt to make the SV more hairy. Ensuring that RBHs exhibit analogous properties of standard BHs, we can include the standard charge, $Q$ and, at most, the angular momentum that, in this study is neglected as we lie on pure spherical symmetry.

To generalize the SV solution, we move from the standard RN metric, defined by $
     A(r)=B(r)= 1-\frac{2M}{r} + \frac{Q^2}{r^2}$,
and, therefore, imposing Eq. \eqref{SV_coordinate}, we compute a class of charged SV solutions\footnote{In classical RN spacetime, the inner (Cauchy) horizon suffers from mass-inflation instabilities. Here, the corresponding regular solution  potentially tames those instabilities.}
\begin{equation}
       A(x)=B(x)= 1-\frac{2M}{\sqrt{x^2 + \ell^2}} + \frac{Q^2}{x^2 + \ell^2}\,.
\end{equation}
As before, we study two fluid models, namely barotropic with $w=0,1$ and with exponential density, say $\rho_0=0.5,1.0\ \text{AU}^{-2}$. We also fix $M=1$, $\eta=0.1$ and set $Q=0.3$, varying  $\ell=0.5,1.5,2.5$, noticing moreover that $\ell\rightarrow0$ again recovers a pure RN solution.

The derivation of the relations from the conservative equations is formally equivalent to the SV case in Sect. \ref{SV_GENERAL_FORMALISM}. The only difference is the metric function $A(x)$ that acquires a new term with charge $Q$. In view of the above results, we can refer to the same analysis of the uncharged SV solutions, even in the charged case.

\subsection{Critical point analysis}
\label{RN_SV_CRITICAL_POINT}

Now, let us focus on the critical point analysis, which involves solving Eq. \eqref{critical_point_eq} as before, in order to determine the variables at $x_c$.

The charged metric modifies the critical-point conditions in Eqs. \eqref{crit_point_conditions} and leads to
\begin{subequations}
    \begin{align}
   \label{Vc_RN} &\, V_c^2 =   \frac{Q^2 - M \sqrt{x^2 + \ell^2}}{3M\sqrt{x^2 + \ell^2}-Q^2 + 2(x^2 + \ell^2)}   \,,\\
   \label{uc_RN} &\, u_c^2 = \frac{M \sqrt{x^2 + \ell^2}-Q^2}{2(x^2 + \ell^2)}\,.
    \end{align}
\end{subequations}
According to the above findings, we can specialize them to each fluid involved into our computation, as below reported.

\subsubsection*{Barotropic fluid}

Equating $u_c^2$ to the barotropic velocity expression, Eq.~\eqref{velocity_expression_barotropic}, leads to
\begin{equation}
  \label{RN_SV_baro_crit}  \left[\frac{\mathcal{C}_4^2}{(1+w)^2} - 1 \right](x^2 + \ell^2) + \frac{3M}{2}\sqrt{x^2 + \ell^2} - \frac{Q^2}{2}=0  \,.
\end{equation}
which has to be solved numerically for each $(\ell,w)$ couple, since the presence of $Q$ does not allow for analytical outcomes.

\subsubsection*{Exponential density profile}

Analogously, for the exponential model we equate Eq. \eqref{uc_RN} with the general counterpart in $x=x_c$ Eq \eqref{exponential_velocity_expression}, and obtain
\begin{equation}
  \label{RN_SV_exp_crit}  \frac{M}{2}(x^2 + \ell^2)^{3/2} - \frac{Q^2}{2}(x^2 + \ell^2) - \frac{\mathcal{C}_3^2}{\rho^2}=0\,.
\end{equation}
As in the first case analysed in Eq. \eqref{crit_point_exp}, we have an expression that cannot be solved analytically, so we will use a numerical tool to solve it.

%
%
\section{Bondi accretion process of uncharged Simpson-Visser spacetime}
\label{RESULTS}

In this section, we present the results of our numerical study of Bondi spherical accretion in the uncharged SV spacetime. Within the two aforementioned fluid treatments, we also consider the following configurations:

\begin{itemize}
  \item[-] A \emph{barotropic} fluid with EoS from Eq. \eqref{equation of state}, using $w=1$ (stiff matter) and $w=0$ (dust matter).
  \item[-] A fluid with an \emph{exponential density profile}, Eq. \eqref{Sofue_density},
  with $\rho_0=0.5,\,1.0\,\mathrm{AU^{-2}}$ and $r_0=10\,\mathrm{AU}$.
\end{itemize}

The mass of the central object is conventionally fixed to  $M=1\,\mathrm{AU}$ and the SV parameter, $\ell$, samples different geometrical solutions allowed as
\begin{itemize}
\item[] $\ell = 0 \rightarrow$, yielding the classical Schwarzschild BH,
\item[] $\ell = 0.5, 1.5 \rightarrow$, representing  intermediate values,  to model RBHs with increasing deviations from Schwarzschild geometry,
\item[] $\ell = 2.5 \rightarrow$, being the critical threshold for horizon formation, resulting into a traversable wormhole geometry.
\end{itemize}
The deviations from Schwarzschild case are treated separately, as confirmed in Subsect. \ref{result_dev_schwarzschild}, where a comparative analysis is performed against the other geometries.

All integration constants, denoted by $\mathcal{C}_i$, are chosen to ensure the existence of a physical sonic (critical) point. Their specific values are listed in the captions of Tables \ref{tab:baro_crit} and \ref{tab:exp_crit}.

We proceed by dividing the discussion in categories: first, we analyze the two RBHs, i.e. $\ell = 0.5, 1.5$ and then, separately, we investigates the wormhole solution\footnote{For the wormhole case ($\ell=2.5$) we plot both $x>0$ and, by mirror symmetry, $x<0$.} ($\ell=2.5$), by plotting together different values of the parameters, therefore $w$ and $\rho_0$, for every fluid model.
In the plots we will see dashed lines for $w=0,\rho_0=0.5$ and continue lines for  $w=1,\rho_0=1$.

The figures for the uncharged solutions are collected in Appendix \ref{appendix}.

\subsection{Regular black hole accretion in Simpson-Visser spacetime}
\label{RBH_SV_results}

We begin by discussing the different RBHs employed in this study and compare them with their classical counterpart, namely the Schwarzschild solution, to assess the impact of the modifications introduced by the SV parameter, namely by $\ell$.

\subsubsection*{Barotropic equation of state}
\label{RBH_SV_barotropic}

We solve the relativistic Bondi equations presented in Sect. \ref{variables_barotropic} for $w = 1$ and $w = 0$. Table \ref{tab:baro_crit} reports the event horizon, $r_{\rm EH}$, the critical radius $x_c$, and the inflow velocity $u_c$ at the critical point for each pair $(\ell, w)$.

We now discuss the distinct trends exhibited by the variables involved in the accretion process.

\begin{figure*}[!h]
  \centering
  \includegraphics[width=0.8\textwidth]{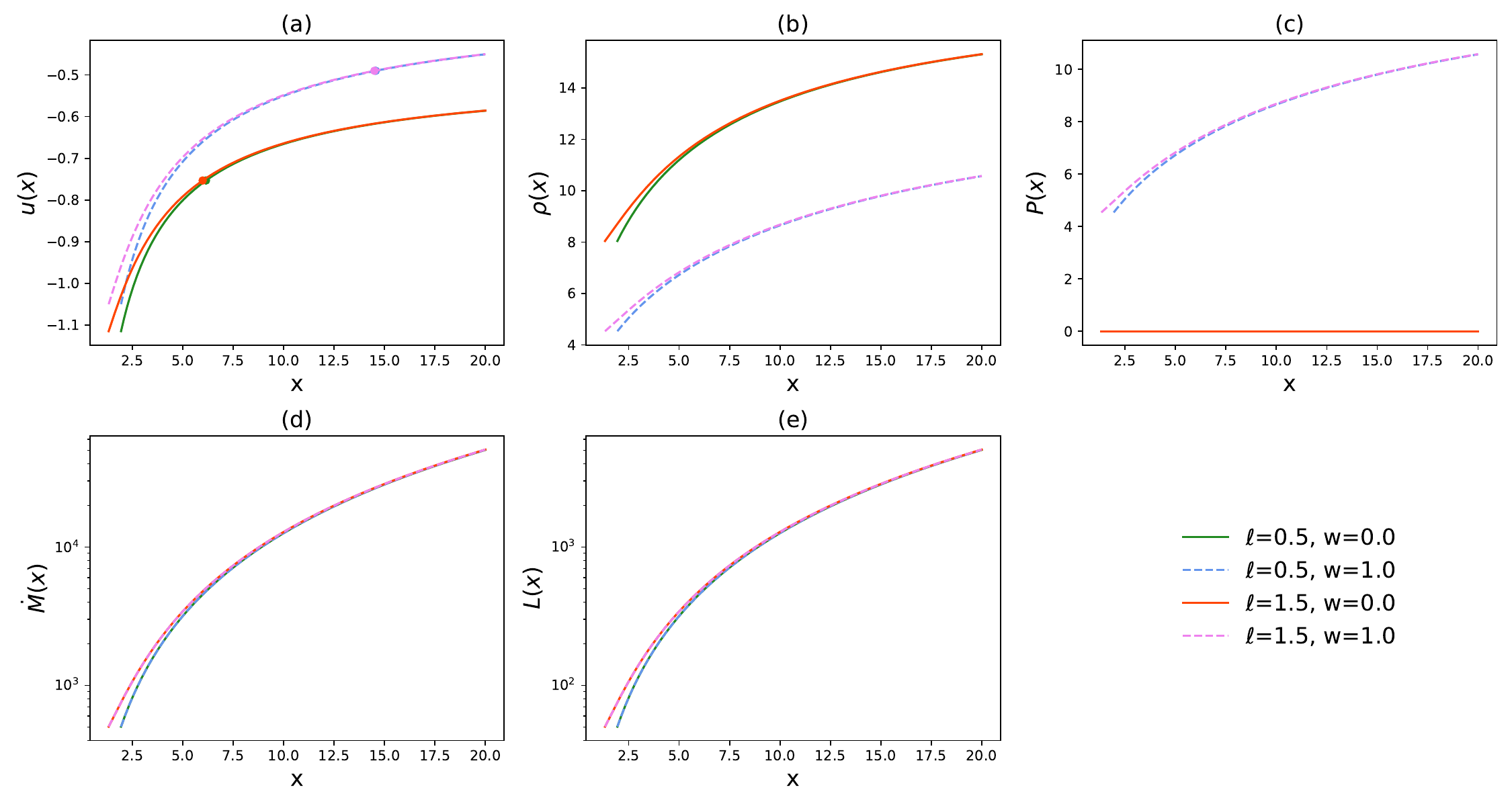}
  \caption{Barotropic EoS ($p=w\rho$) profiles \emph{w.r.t.} $x/M$: (a) radial velocity $u(x)$, (b) density $\rho(x)$, (c) pressure $P(x)$, (d) accretion rate $\dot M(x)$, and (e) luminosity $L(x)$, for $\ell=0.5,1.5$ and $w=0,1$. Critical radii are marked with filled circles in panel (a).}
  \label{fig:RBH_SV_barotropic}
\end{figure*}

Let us first examine Fig. \ref{fig:RBH_SV_barotropic}. In panel (a), we plot the fluid velocities as a function of $x/M$. The solutions are split into two groups, corresponding to the two types of matter considered: \emph{stiff matter} ($w = 1$) and \emph{dust}  ($w = 0$). As observed, the trends of the two fluids are quite similar. The solutions for each $\ell$ are nearly superposed in the inner region and asymptotically approach zero, although they decrease at different rates but almost reach a superposition near the event horizon.

Afterwards, we analyze the behavior of the critical points for each solution. For stiff matter, the flow becomes supersonic at a larger radius, farther from the event horizon. Notably, for dust, the critical point analysis reveals the absence of physical critical points using the same integration constant $\mathcal{C}_4$ as in the stiff matter case. Hence, adopting a different value, as shown in Table \ref{tab:baro_crit}, allows to adjust the corresponding results. Accordingly, these modified values are smaller than those for stiff matter, indicating that the transition to the supersonic regime occurs deeper within the accretion disk, as expected from the nature of stiff matter and dust.

The corresponding density and pressure, shown in panels (b) and (c), are constrained by Eq. \eqref{equation of state} and,  thus, follow similar trends. Since dust is essentially pressureless, one set of solutions exhibits a flat, null pressure profile. Conversely, stiff matter presents increasing pressure and density profiles in the outer region. Both variables again separate into two distinct groups based on the value of $w$.

Importantly, we observe these increasing trends in density and pressure both physically and numerically. By truncating the spacetime at $40~\mathrm{AU}$\footnote{This cutoff was used solely to verify the asymptotic behavior; for simulating accretion variables, we stopped at $20~\mathrm{AU}$.}, we verified that these quantities reach a plateau and do not diverge. This conclusion also applies to the accretion rate and luminosity discussed below, as they depend on the density.

Panels (d) and (e) present the accretion rate and disk luminosity for the barotropic fluid, both plotted on a logarithmic scale. As expected, these quantities increase in the outer regions of the accretion flow due to their dependence on the density profile. Although density and pressure scale differently depending on $w$, the expressions for accretion rate, $\dot{M}$, and luminosity, $L$, yield comparable results. The $w = 0$ curves may appear missing, but are superimposed on the $w = 1$ curves.

Thus, we observe a similar increase in both accretion rate and luminosity in the outer disk regions for all solutions. Dust and stiff matter cases exhibit overlapping behavior, despite differences in their respective density profiles.

Moreover, since luminosity differs from the mass accretion rate only by the efficiency parameter $\eta = 0.1$ (see Eq. \eqref{bondi_luminosity}), the luminosity values in panel (e) are approximately one order of magnitude lower than $\dot{M}$ at the same $x$.

\begin{figure*}[!h]
  \centering
  \includegraphics[width=0.8\textwidth]{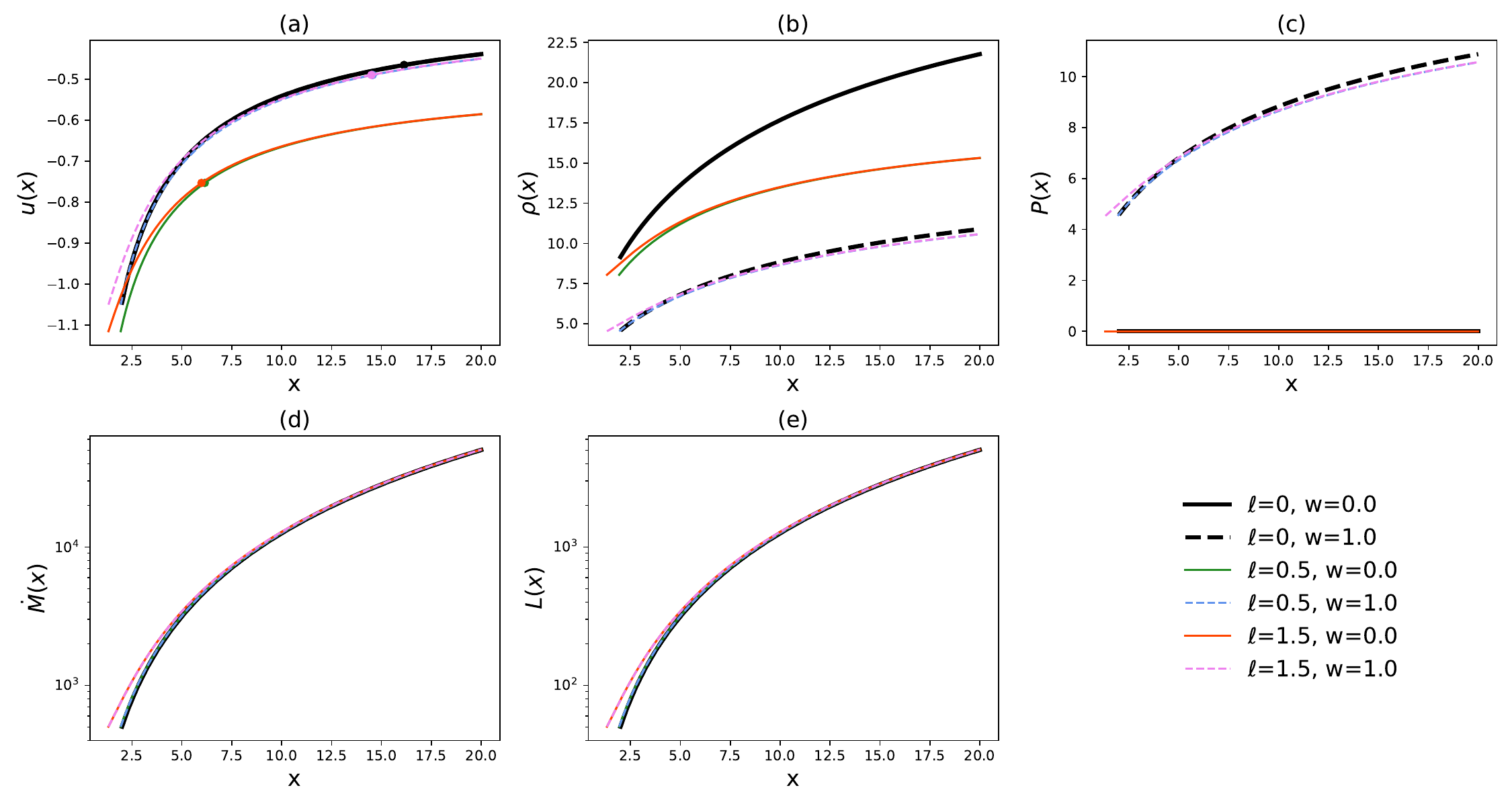}
  \caption{Comparison between Schwarzschild ($\ell=0$, black lines) and RBHs solutions for a barotropic EoS ($p=w\rho$) profiles \emph{w.r.t.} $x/M$: (a) radial velocity $u(x)$, (b) density $\rho(x)$, (c) pressure $P(x)$, (d) accretion rate $\dot M(x)$, and (e) luminosity $L(x)$, for $\ell=0.5,1.5$ and $w=0,1$. Critical radii are marked with filled circles in panel (a).}
  \label{fig:RBH_SV_barotropic_comparison}
\end{figure*}

Quite importantly, we can now  compare the RBH solutions with the Schwarzschild BH, as shown in Fig. \ref{fig:RBH_SV_barotropic_comparison}, where the Schwarzschild solution is plotted  by black lines, i.e., solid and dashed to distinguish different $w$ values. The corresponding numerical results for the critical point analysis are reported in Table \ref{tab:baro_crit}, along with all the parameters and constants used.

In panel (a), the Schwarzschild solutions closely match the RBH curves, particularly those with $\ell = 0.5$ for both values of $w$. Slight deviations occur in the $\ell = 1.5$ case, especially near the event horizon.

Remarkably, from Eq. \eqref{exponential_velocity_expression}, it appears clear that the dependence on $w$ is very weak and, so, the two classical cases are superimposed. The critical radii $x_c$ for the various Schwarzschild plots are slightly larger, albeit remaining of the same order as those of our RBHs scenarios.

Similarly, for $w = 1$, the density profiles are almost indistinguishable between RBHs and Schwarzschild cases, indicating that small values of $\ell$ do not significantly alter the fluid structure. For $w = 0$, however, the Schwarzschild solution shows a marked increase in density at large radii, deviating from the RBH behavior. Pressure profiles, on the other hand, exhibit good agreement across all models and the classical case.

Accretion rate and luminosity, reported in panels (d) and (e),  also show strong agreement, with minor enhancements in RBH models as $\ell$ increases, particularly for wormhole-like solutions. These differences, however, are subtle and mainly appear near the horizon.

\begin{table*}[!t]
  \centering
  \renewcommand{\arraystretch}{1.2}
    \begin{tabular}{l@{\hspace{1cm}}c@{\hspace{0.5cm}}c@{\hspace{0.5cm}}c@{\hspace{0.5cm}}c}

    \hline
    \multicolumn{4}{c}{\textbf{Critical Point Analysis: Barotropic Fluid}} \\
    \hline
    $(\ell,w)$ & $r_{\rm EH}$ & $x_c$ & $u_c$ & $\mathcal{C}_4$\\
    \hline
    (0,\   1)  & 2.00   &  16.13  &  -0.47  &  2.09    \\
    (0,\   0)  & 2.00   &  16.13  &  -0.47  &  1.05    \\
    (0.5,\ 1)  & 1.94   &  14.57  &  -0.49  &  2.10    \\
    (0.5,\ 0)  & 1.94   &  6.15   &  -0.75  &  1.12    \\
    (1.5,\ 1)  & 1.32   &  14.50  &  -0.49  &  2.10    \\
    (1.5,\ 0)  & 1.32   &  5.98   &  -0.75  &  1.12    \\
    (2.5,\ 1)  &   -    &  14.36  &  -0.49  &  2.10    \\
    (2.5,\ 0)  &   -    &  5.64   &  -0.75  &  1.12    \\
    \hline
  \end{tabular}
  \caption{Event horizon, sonic radius, and flow speed at the sonic point for each $(\ell,w)$ in the barotropic model. For the analysis, we used the following values for the integration constants: $\mathcal{C}_1=10, \ \mathcal{C}_3=1.9$ and the parameters $M=1\mathrm{AU},\ \eta=0.1$.}
  \label{tab:baro_crit}
\end{table*}


\subsubsection*{Exponential density profile}
\label{RBH_SV_exponential}

\begin{figure*}[!t]
  \centering
  \includegraphics[width=0.8\textwidth]{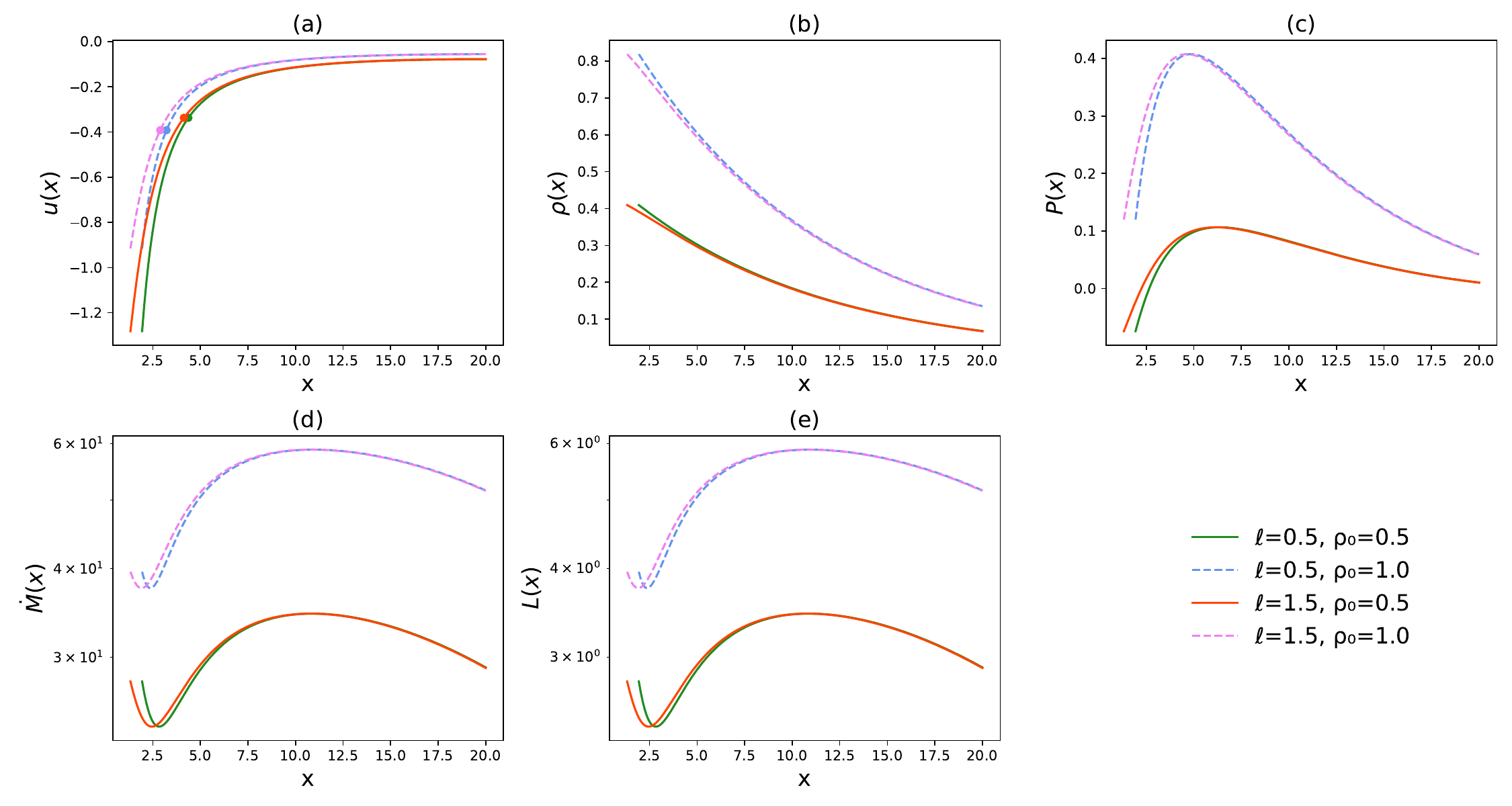}
  \caption{Exponential density profile model \emph{w.r.t.} $x/M$: (a) radial velocity $u(x)$, (b) density $\rho(x)$, (c) pressure $P(x)$, (d) accretion rate $\dot M(x)$, and (e) luminosity $L(x)$, for $\ell=0.5,1.5$ and $\rho_0=0.5,1.0\mathrm{AU}^{-2}$. The critical points are marked in the velocity panel.}
  \label{fig:RBH_exp_SV}
\end{figure*}

The exponential case is illustrated in Fig. \ref{fig:RBH_exp_SV}, where the primary influence on the density profile is due to the parameter $\rho_0$, rather than by $w$. As anticipated, we consider two values for this parameter: $0.5$ and $1~\mathrm{AU}^{-2}$. The choice of these values comes from the fact that for lower values of $\rho_0$, no physical critical points could be found with our setup. The constants and parameters used in this analysis are listed in Table \ref{tab:exp_crit}.

In panel (a), the inflow velocity shows a trend similar to that of the barotropic fluid, but with notable differences in the inner regions near the event horizon. The velocity curves for the two $\rho_0$ values overlap in the outer region, unlike in the barotropic case. However, closer to the horizon, the behavior diverges slightly, forming two distinct groups depending on the value of $\rho_0$.

The critical points lie closer to the event horizon compared with the barotropic case. This shift may result from the different density profiles, parameter choices, and integration constants. In particular, the critical radii associated with $\rho_0 = 0.5~\mathrm{AU}^{-2}$ are systematically larger than those corresponding to $\rho_0 = 1~\mathrm{AU}^{-2}$, consistent with the dependence in Eq. \eqref{crit_point_exp}.

Panel (b) shows the exponential density profiles, clearly forming two separate groups due to the variation in $\rho_0$, while the value of $r_0$ is held constant. These profiles asymptotically decrease as one moves away from the accretion disk.

Panel (c) presents the pressure profiles, which are similarly grouped by $\rho_0$. While the pressure roughly mirrors the density trends, a distinct drop in pressure is observed as the flow approaches the event horizon.

Panels (d) and (e) display the accretion rate and luminosity, respectively, both on logarithmic scales. The two plots differ only by the inclusion of the efficiency parameter $\eta$ in the luminosity formula. In contrast to the barotropic case, where all solutions nearly overlap, here the solutions are distinct, based on $\rho_0$. The overall behavior also differs: starting from small $x$, namely near the horizon, both $\dot{M}$ and $L$ initially decrease rapidly, and as $x$ increases, both trends grow again before experiencing a damped decrease. This behavior is observed across all configurations, independently of $\rho_0$, with the only difference consisting of  the maximum value attained before the drop, which approaches zero as $x \rightarrow \infty$.

\begin{figure*}[!t]
  \centering
  \includegraphics[width=0.8\textwidth]{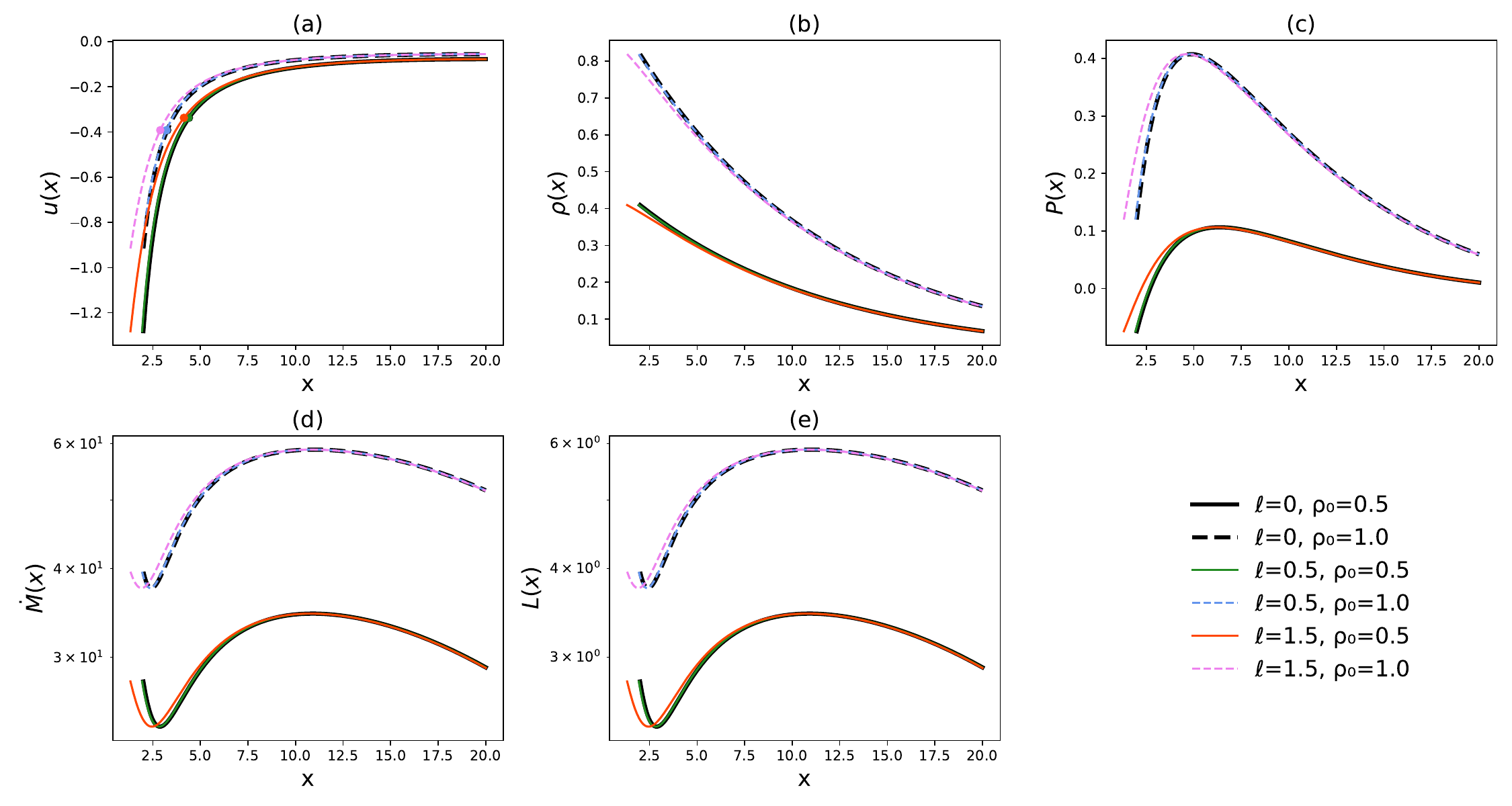}
  \caption{Comparison between Schwarzschild ($\ell=0$, black lines) and RBHs solutions for an exponential density profile model \emph{w.r.t.} $x/M$: (a) radial velocity $u(x)$, (b) density $\rho(x)$, (c) pressure $P(x)$, (d) accretion rate $\dot M(x)$, and (e) luminosity $L(x)$, for $\ell=0.5,1.5$ and $\rho_0=0.5,1.0\mathrm{AU}^{-2}$. The critical points are marked in the velocity panel.}
  \label{fig:RBH_exp_SV_comparison}
\end{figure*}

Remarkably, we now compare the exponential profile solutions with the Schwarzschild BH, as shown in Fig. \ref{fig:RBH_exp_SV_comparison}. The classical solution, depicted by black lines, closely resembles the RBH cases with small $\ell$.

Inflow velocity, density, and pressure profiles, shown in panels (a), (b), and (c), respectively, begin to diverge near the center. The critical radius increases by up to $\sim6$ for moderate values of $\ell$. Unlike the barotropic model, the exponential profile is less sensitive to both the value of $\rho_0$ and the underlying geometry. Indeed, the velocity profiles exhibit negligible differences in magnitude and are nearly indistinguishable.

In panel (d), the Schwarzschild mass accretion rate follows the same trend as the SV RBH solutions, with all curves nearly superimposed and showing a slightly smoother decrease near the horizon. This reflects the influence of spacetime regularization on the fluid dynamics, particularly when paired with non-trivial density profiles.

Finally, panel (e) shows the luminosity profiles,  mirroring the accretion rate behavior, confirming that the Schwarzschild solution is indistinguishable for small $\ell$.

In all plots, comparing with the Schwarzschild solution, the most significant discrepancies appear near the event horizon. In these regions, the classical solution tends to exhibit growth or decay earlier than the RBH models, illustrating the distinctive response of regularized spacetimes.

\begin{table*}[!t]
  \centering
  \renewcommand{\arraystretch}{1.2}
  \begin{tabular}{l@{\hspace{1cm}}c@{\hspace{0.5cm}}c@{\hspace{0.3cm}}c@{\hspace{0.3cm}}c}
    \hline
    \multicolumn{4}{c}{\textbf{Critical Point Analysis: Exponential Profile}} \\
    \hline
    $(\ell,\rho_0)$ & $r_{\rm EH}$ & $x_c$ & $u_c$ & $\mathcal{C}_3$\\
    \hline
    (0,\   0.5)  & 2.00    & 4.397     & -0.34   & 2.10  \\
    (0,\     1)  & 2.00    & 3.256     & -0.39   & 3.00  \\
    (0.5,\ 0.5)  & 1.94     & 4.37     & -0.34   & 2.10  \\
    (0.5,\   1)  & 1.94     & 3.22     & -0.39   & 3.00  \\
    (1.5,\ 0.5)  & 1.32     & 4.13     & -0.34   & 2.10  \\
    (1.5,\   1)  & 1.32     & 2.90     & -0.39   & 3.00  \\
    (2.5,\ 0.5)  &  -       & 4.40     & -0.34   & 2.10  \\
    (2.5,\   1)  &  -       & 3.26     & -0.39   & 3.00  \\
    \hline
  \end{tabular}
  \caption{Event horizon, sonic radius, and flow speed at the sonic point for each $(\ell,\rho_0)$ in the exponential profile model. For the analysis, we used the following values for the integration constants: $\mathcal{C}_=10$,  $\mathcal{C}_4=2.1$ and the parameters $M=1\mathrm{AU},\ r_0 = 10 \mathrm{AU},\ \eta=0.1$.}
  \label{tab:exp_crit}
\end{table*}
%

\subsection{Wormholes accretion in Simpson-Visser spacetime}
\label{WH_SV_results}

We now present the results for the wormhole configuration, corresponding to $\ell = 2.5$, for both fluid models, in analogy with the RBH case. As anticipated previously, we extend the profiles to negative values of $x$ by imposing symmetry of the solution across the throat.

Again, the two configurations for barotropic and exponential fluids are reported below.

%
\subsubsection*{Barotropic case}
\label{WH_SV_barotropic}

\begin{figure*}[!t]
  \centering
  \includegraphics[width=0.8\textwidth]{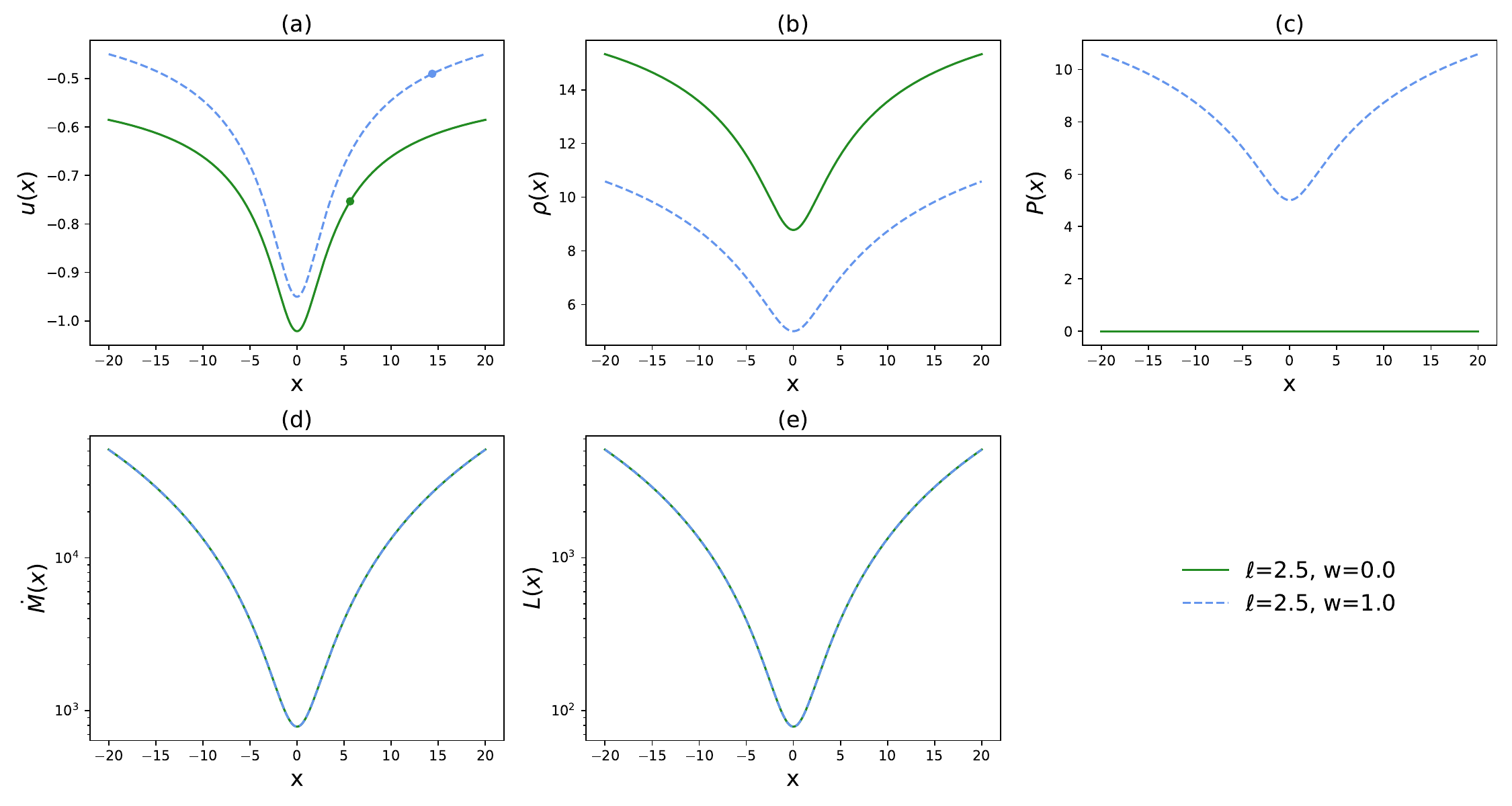}
  \caption{Barotropic EoS ($p=w\rho$) profiles \emph{w.r.t.} $x/M$: (a) radial velocity $u(x)$, (b) density $\rho(x)$, (c) pressure $P(x)$, (d) accretion rate $\dot M(x)$, and (e) luminosity $L(x)$, for $\ell=2.5$ and $w=0,1$. Critical radii are marked with filled circles in panel (a).}
  \label{fig:WH_SV_barotropic}
\end{figure*}

Fig. \ref{fig:WH_SV_barotropic} shows the barotropic solutions in the wormhole regime. In panel (a), the inflow velocity behaves similarly to the RBH case. The two profiles, corresponding to $w=1$ and $w=0$, tend to get closer but without overlapping near the throat, $x=0$, and diverge slightly in the outer regions, where the stiffer fluid reaches a higher asymptotic speed.

The critical points for wormholes are of the same order of magnitude as those in the RBH cases, differing only in decimal precision. For stiff matter, the critical point lies farther from the throat, while for dust matter, it is located closer to the center.

Panels (b) and (c) display the density and pressure profiles. As dictated by the equation of state, the two variables follow the same general trend: both increase moving outward and decrease smoothly near the throat. Unlike in BH configurations, the absence of an horizon ensures the continuity of the solution, including the point $x = 0$.

The accretion rate and luminosity in panels (d) and (e) reflect the behavior of the density and pressure. Both quantities grow with $x$ in the outer region, then decrease towards the throat. Their magnitudes are comparable to those in the RBHs case, but with noticeably smoother transitions near $x = 0$.

\begin{figure*}[!t]
  \centering
  \includegraphics[width=0.8\textwidth]{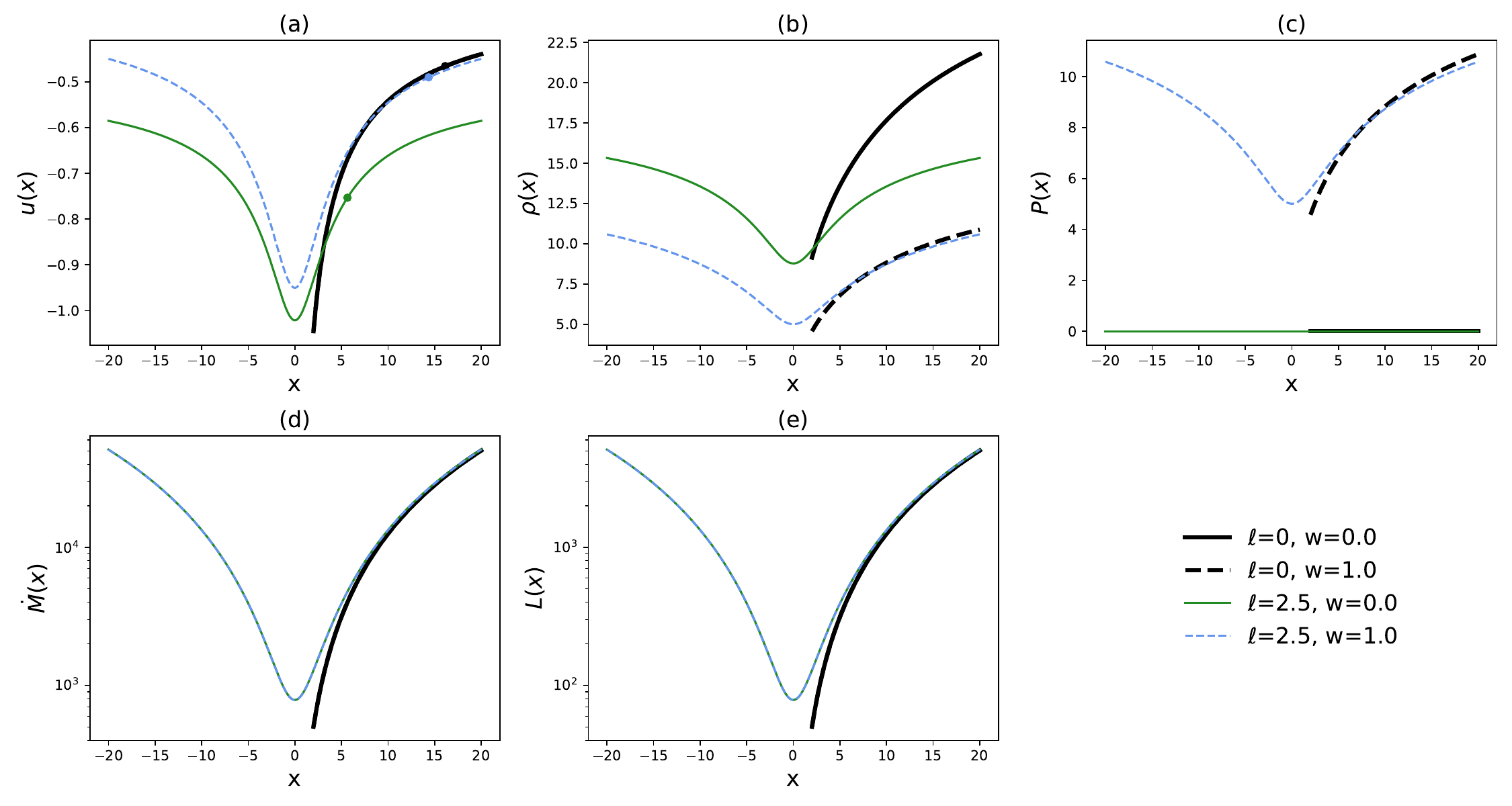}
  \caption{Comparison between Schwarzschild ($\ell=0$, black lines) and RBHs solutions for a barotropic EoS ($p=w\rho$) profiles \emph{w.r.t.} $x/M$: (a) radial velocity $u(x)$, (b) density $\rho(x)$, (c) pressure $P(x)$, (d) accretion rate $\dot M(x)$, and (e) luminosity $L(x)$, for $\ell=2.5$ and $w=0,1$. Critical radii are marked with filled circles in panel (a).}
  \label{fig:WH_SV_barotropic_comparison}
\end{figure*}

Analogously to what computed before, we now compare the wormhole and Schwarzschild solutions in Fig. \ref{fig:WH_SV_barotropic_comparison}. In panel (a), the radial velocity of the Schwarzschild solution rises steeply near the horizon, whereas the wormhole solutions feature a smoother gradient. Notably, the Schwarzschild profiles nearly coincide with the wormhole ones in the dust case, $w = 0$, albeit not for stiff matter. The critical radius $x_c$ in the wormhole configuration is slightly larger, but still of the same order.

In panel (b), the density profiles for dust match closely, while in the stiff matter case the Schwarzschild profile shows more rapid growth in the outer region. Panel (c) reveals good agreement in pressure, across all cases, with minor differences near the center.

Finally, panels (d) and (e) show the accretion rate and luminosity. These remain similar in magnitude between the Schwarzschild and wormhole configurations, with minor enhancements in the former near the horizon. Differences remain modest and appear primarily near $x = 0$.

%
\subsubsection{Exponential density profile}
\label{WH_SV_exponential}

\begin{figure*}[!t]
  \centering
  \includegraphics[width=0.8\textwidth]{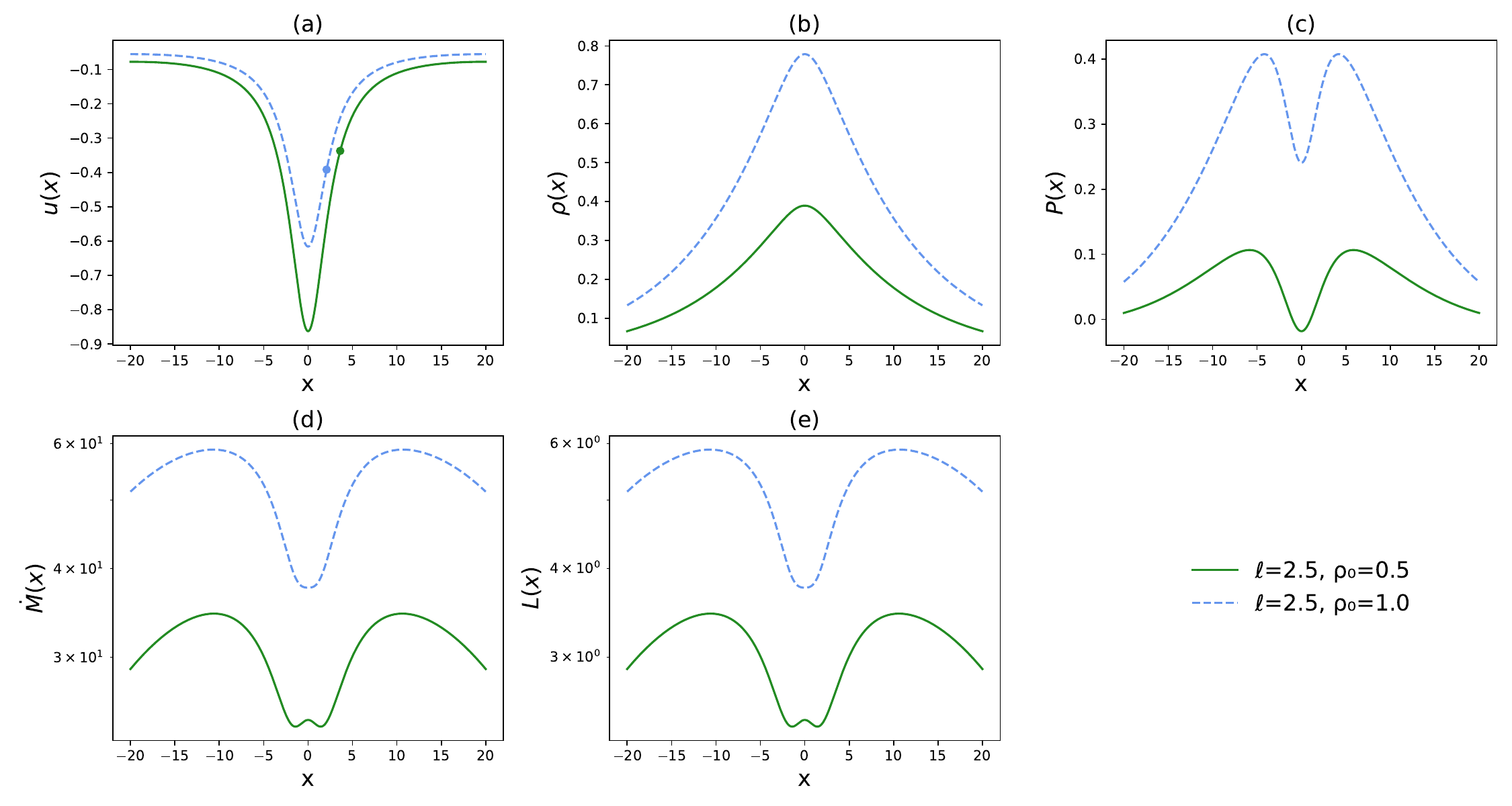}
  \caption{Exponential density profile model \emph{w.r.t.} $x/M$: (a) radial velocity $u(x)$, (b) density $\rho(x)$, (c) pressure $P(x)$, (d) accretion rate $\dot M(x)$, and (e) luminosity $L(x)$, for $\ell=2.5$ and $\rho_0=0.5,1.0\mathrm{AU}^{-2}$. The critical points are marked in the velocity panel.}
  \label{fig:WH_exp_SV}
\end{figure*}

The wormhole case with an exponential density profile is shown in Fig. \ref{fig:WH_exp_SV}. The inflow velocity, panel (a), is again split into two sets of curves depending on the value of $\rho_0$. The velocity decreases outward and remains symmetrical with respect to the throat. Near $x = 0$, the falloff is smoother than in BH cases, especially for higher density, characterized by $\rho_0 = 1$, indicating a shallower potential well.

The critical points in this setup are comparable to those of other solutions, with slightly larger critical radii observed for $\rho_0 = 0.5$.

As seen in panels (b) and (c), density and pressure decrease smoothly toward the outer regions and rise towards the center, similar to the behavior in the barotropic model. Both quantities remain continuous across the throat, and no discontinuities are observed.

The accretion rate and luminosity in panels (d) and (e) mirror those found for RBHs. Both observables exhibit a gradual decline at large $x$, a central minimum near $x = 0$, and a mild secondary peak for $\rho_0 = 0.5$ before reaching the throat.

Overall, the most notable differences from other solutions occur near the throat, where regularity of the geometry moderates the fluid gradients.

\begin{figure*}[!t]
  \centering
  \includegraphics[width=0.8\textwidth]{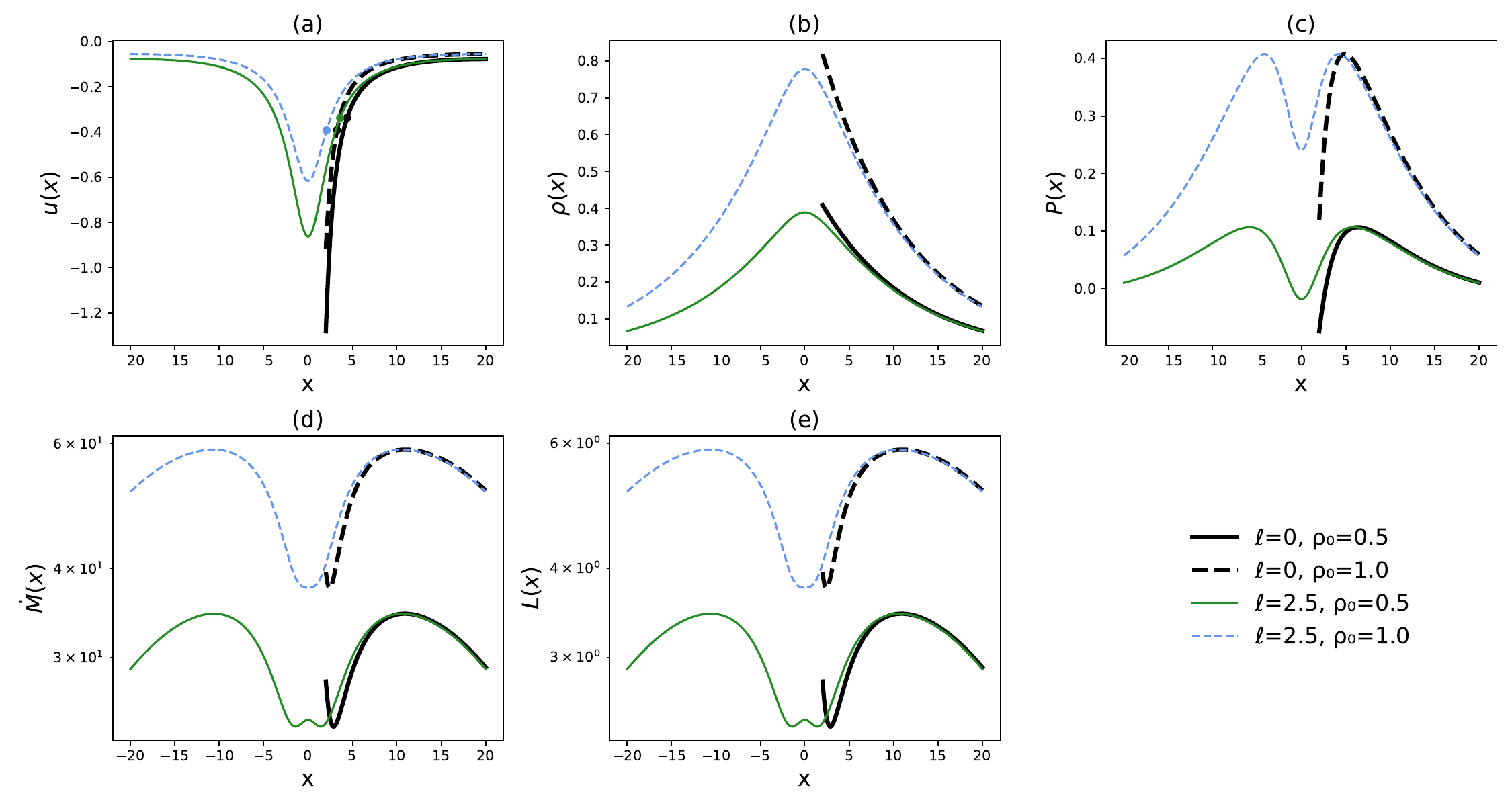}
  \caption{Comparison between Schwarzschild ($\ell=0$, black lines) and RBHs solutions for an exponential density profile model \emph{w.r.t.} $x/M$: (a) radial velocity $u(x)$, (b) density $\rho(x)$, (c) pressure $P(x)$, (d) accretion rate $\dot M(x)$, and (e) luminosity $L(x)$, for $\ell=2.5$ and $\rho_0=0.5,1.0\mathrm{AU}^{-2}$. The critical points are marked in the velocity panel.}
  \label{fig:WH_exp_SV_comparison}
\end{figure*}

The comparison with the Schwarzschild solution is shown in Fig. \ref{fig:WH_exp_SV_comparison}. In panel (a), the radial velocity again increases steeply near the Schwarzschild horizon, while the wormhole solution shows a smoother decline, especially for $\rho_0 = 1$. The Schwarzschild profiles match closely with the wormhole ones in the dust matter case, albeit deviating significantly for stiff matter.

Panel (b) shows good agreement in the density profiles at large radii, with notable divergence near the center, where wormhole solutions dip and Schwarzschild continues to rise. The pressure profiles in panel (c) follow a similar pattern: excellent overlap at large $x$ but with the Schwarzschild pressure decreasing more steeply in the inner region.

In panels (d) and (e), the accretion rate and luminosity again align closely for both geometries, with small enhancements for Schwarzschild near the horizon. The wormhole solution with $\rho_0 = 0.5$ shows a minor central peak, but it is far less pronounced than the classical growth seen in the Schwarzschild model.

%
\subsubsection{Quantifying deviations from the Schwarzschild solution}
\label{result_dev_schwarzschild}

Above, our analyses focused on discrepancies among solutions and, in particular, using the Schwarzschild BH.

It appears convenient to  quantify how much the SV scenarios differ from the simplest spherically-symmetric solution, namely the Schwarzschild metric, in order to shed light toward possible future observations that can \emph{disentangle} the latter standard solution from the SV spacetimes.

Hence, to quantify the deviations from the classical Schwarzschild BH, introduced by employing a non-zero $\ell$, we define the relative percentage deviation as
\begin{equation}
\label{percentage_deviation}
\Delta Q(\ell) = \frac{|Q(\ell) - Q(0)|}{Q(0)} \times 100\%, \quad Q \in \{x_{\rm EH},\, x_c,\, u_c\}.
\end{equation}

Table \ref{tab:percent_dev} reports the resulting percentage deviations in the horizon location $x_{\rm EH}$, critical radius $x_c$, and critical inflow velocity $u_c$ for all cases with $\ell > 0$, for both barotropic and exponential fluid models.

\begin{table*}[!t]
  \centering
  \renewcommand{\arraystretch}{1.2}
  \begin{tabular}{c c c c c c}
    \hline
    Model & Param & $\ell$ & $\Delta x_{\rm EH}[\%]$ & $\Delta x_c[\%]$ & $\Delta u_c[\%]$ \\
    \hline
    \multirow{3}{*}{Barotropic EoS, $w=0$}
      &                 & 0.5   & $3.18$        & $0.05$  & $0$     \\
      &                 & 1.5   & $33.86$       & $0.43$  & $0$     \\
      &                 & 2.5   & ---           & $1.21$  & $0$     \\
    \hline
    \multirow{3}{*}{Barotropic EoS, $w=1$}
      &                 & 0.5   & $3.18$        & $0.05$  & $0$     \\
      &                 & 1.5   & $33.86$       & $0.43$  & $0$     \\
      &                 & 2.5   & ---           & $1.21$  & $0$     \\
    \hline
    \multirow{3}{*}{Exponential, $\rho_0=0.5$}
      &                 & 0.5   & $3.18$        & $0.65$  & $0$     \\
      &                 & 1.5   & $33.86$       & $6.00$  & $0$     \\
      &                 & 2.5   & ---           & $17.74$ & $0$     \\
    \hline
    \multirow{3}{*}{Exponential, $\rho_0=1.0$}
      &                 & 0.5   & $3.18$        & $1.19$  & $0$     \\
      &                 & 1.5   & $33.86$       & $11.24$ & $0$     \\
      &                 & 2.5   & ---           & $35.93$ & $0$     \\
    \hline
  \end{tabular}
  \caption{Percentage deviations of the horizon coordinate $x_{\rm EH}$, sonic radius $x_c$, and critical velocity $u_c$ for $\ell > 0$, relative to the Schwarzschild solution. Entries marked “$---$” indicate the absence of an event horizon in the wormhole regime ($\ell = 2.5$).}
  \label{tab:percent_dev}
\end{table*}

From the outcomes, we observe that for moderate regularization scales, i.e., $\ell = 0.5$, the deviation in the event horizon is roughly $3\%$, whereas the critical radius shifts by less than $1\%$. These variations are small, suggesting limited impact on the accretion dynamics.

For larger values, such as $\ell = 1.5$, deviations in $x_c$ become more noticeable: around $6\%$ for the exponential model with $\rho_0 = 0.5$ and up to $11\%$ for $\rho_0 = 1.0$. The barotropic model remains less sensitive, with changes in $x_c$ still well below $1\%$.

In the wormhole regime, $\ell = 2.5$, we observe substantial differences in the exponential model, particularly with $\rho_0 = 1.0$, where $x_c$ deviates by nearly $36\%$. Nonetheless, for barotropic fluids, the shift remains modest, namely $\sim 1\%$, indicating that the SV deformation affects sonic transitions weakly unless paired with strong density gradients.

In all the configurations, the critical fluid velocity $u_c$ remains constant to within numerical precision, further confirming the robustness of the accretion process with respect to regularization effects.

\section{Bondi accretion for charged Simpson-Visser spacetime}
\label{RN_RESULTS}

Our previous findings are based on using SV solutions, involving only the mass as free parameter besides $\ell$.

Quite remarkably, especially for astrophysical cases of interest it would be interesting to work SV solutions out by adding electric charge, $Q$, in analogy to the RN solution.

Accordingly, we here analyze the impact of $Q$ to  accretion profiles in the \emph{charged SV spacetime}, comparing each regularized configuration, $\ell > 0$, to the classical RN baseline, $\ell = 0,\ Q\neq0$.

Notice that the role played by $Q$ is that of a physical charge, thus quite different than the topological charges present in several versions of RBHs, such as the Bardeen solution.

Immediately, the computed values of the event horizon $x_{\rm EH}$, critical radius $x_c$, and critical inflow velocity $u_c$ for each fluid model, that is, for each pair $(\ell,w)$ and $(\ell,\rho_0)$, are reported in Tables \ref{tab:barotropic_RN} and \ref{tab:Exp_RN}, respectively.

Afterwards, we explored several values of the charge $Q$ satisfying $Q \leq M$, ultimately fixing $Q = 0.3$\footnote{We also tested values of $Q$ close to $M$ and significantly smaller values. The results showed negligible qualitative differences compared to $Q = 0.3$.}.

The figures for the charged solutions are collected in Appendix \ref{appendix}.

\subsection{Regular black hole accretion}
\label{RBH_RN_results}

We begin with the charged RBHs configurations corresponding to $\ell = 0.5$ and $1.5$, with $M = 1$ and $Q = 0.3$.

\subsubsection*{Barotropic equation of state}
\label{RBH_RN_barotropic}

In the charged SV spacetime, the barotropic accretion profile preserves the same analytical structure as in the uncharged case, with the metric function now including the electric charge $Q = 0.3M$. As shown in Tab. \ref{tab:barotropic_RN}, the presence of charge induces small shifts in the positions of the event horizon $x_{\rm EH}$ and the sonic radius $x_c$, while the critical velocity $u_c$ remains essentially unchanged.

\begin{figure*}[!t]
  \centering
  \includegraphics[width=0.8\textwidth]{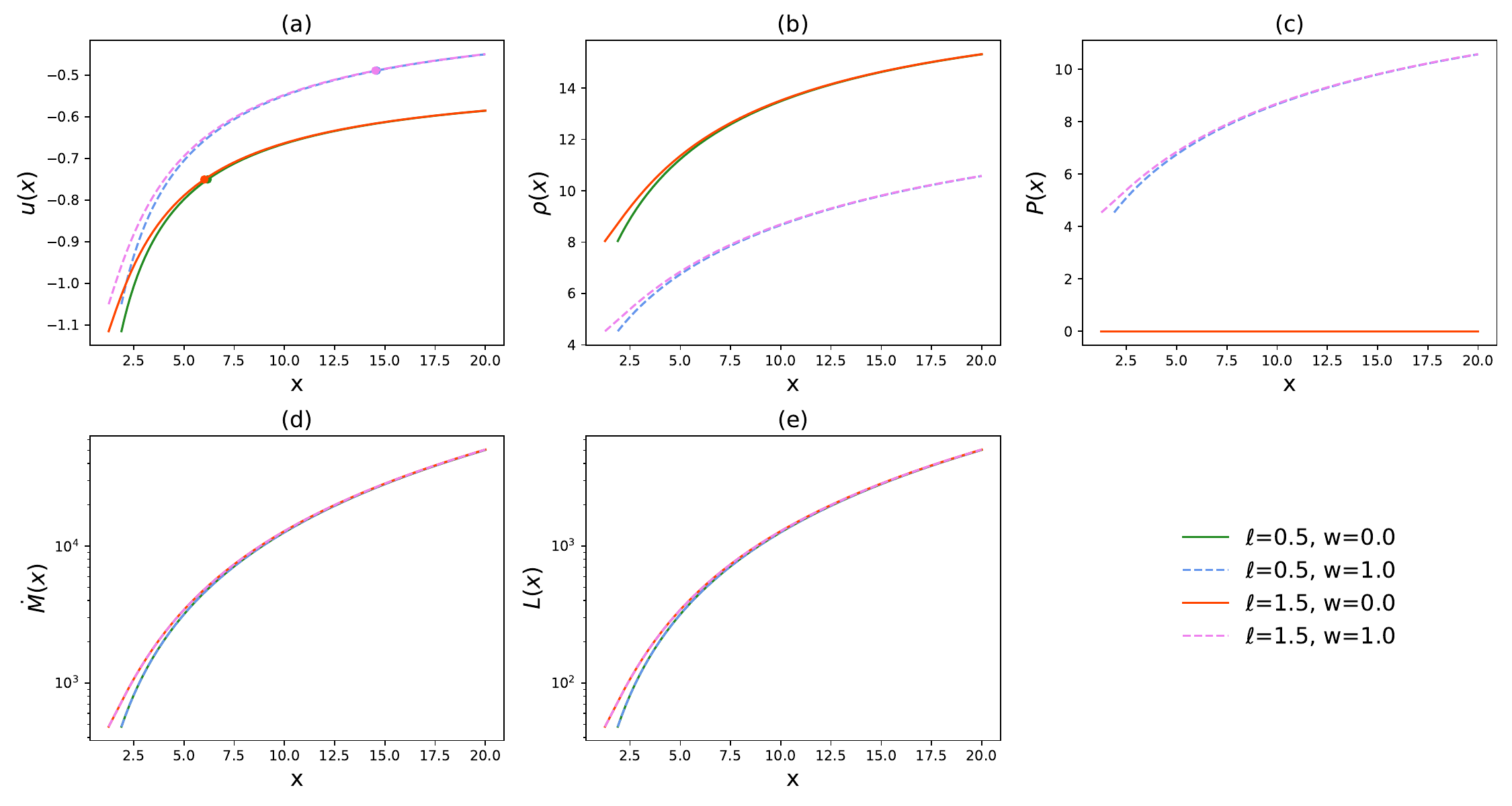}
  \caption{Barotropic EoS ($p=w\rho$) profiles \emph{w.r.t.} $x/M$: (a) radial velocity $u(x)$, (b) density $\rho(x)$, (c) pressure $P(x)$, (d) accretion rate $\dot M(x)$, and (e) luminosity $L(x)$, for $\ell=0.5,1.5$, $w=0,1$ and $Q=0.3$. Critical radii are marked with filled circles in panel (a).}
  \label{fig:RBH_RN_barotropic}
\end{figure*}

Figure \ref{fig:RBH_RN_barotropic} displays the radial profiles of velocity, density, pressure, mass accretion rate, and luminosity. Panel (a) shows that near the horizon, all velocity profiles nearly overlap, with a slight suppression for $\ell = 0.5$. In the outer region, the solutions split into two distinct branches corresponding to $w = 0$ and $w = 1$. The stiff matter solutions continue to  exhibite critical points farther from the center than those for dust matter.

Density and pressure profiles, shown in panels (b) and (c), follow the same trends observed for neutral RBHs. Stiff matter exhibits higher densities and pressures overall. The electric charge slightly reduces the central density due to weakened gravitational attraction. For $w = 0$, the pressure is identically zero  and the density profile may flatten near the center.

Accretion rate and luminosity, shown in panels (d) and (e), closely follow the behavior observed in the neutral case. Variations due to charge are minimal, and the magnitude and qualitative shape of the profiles remain unchanged.

\begin{figure*}[!t]
  \centering
  \includegraphics[width=0.8\textwidth]{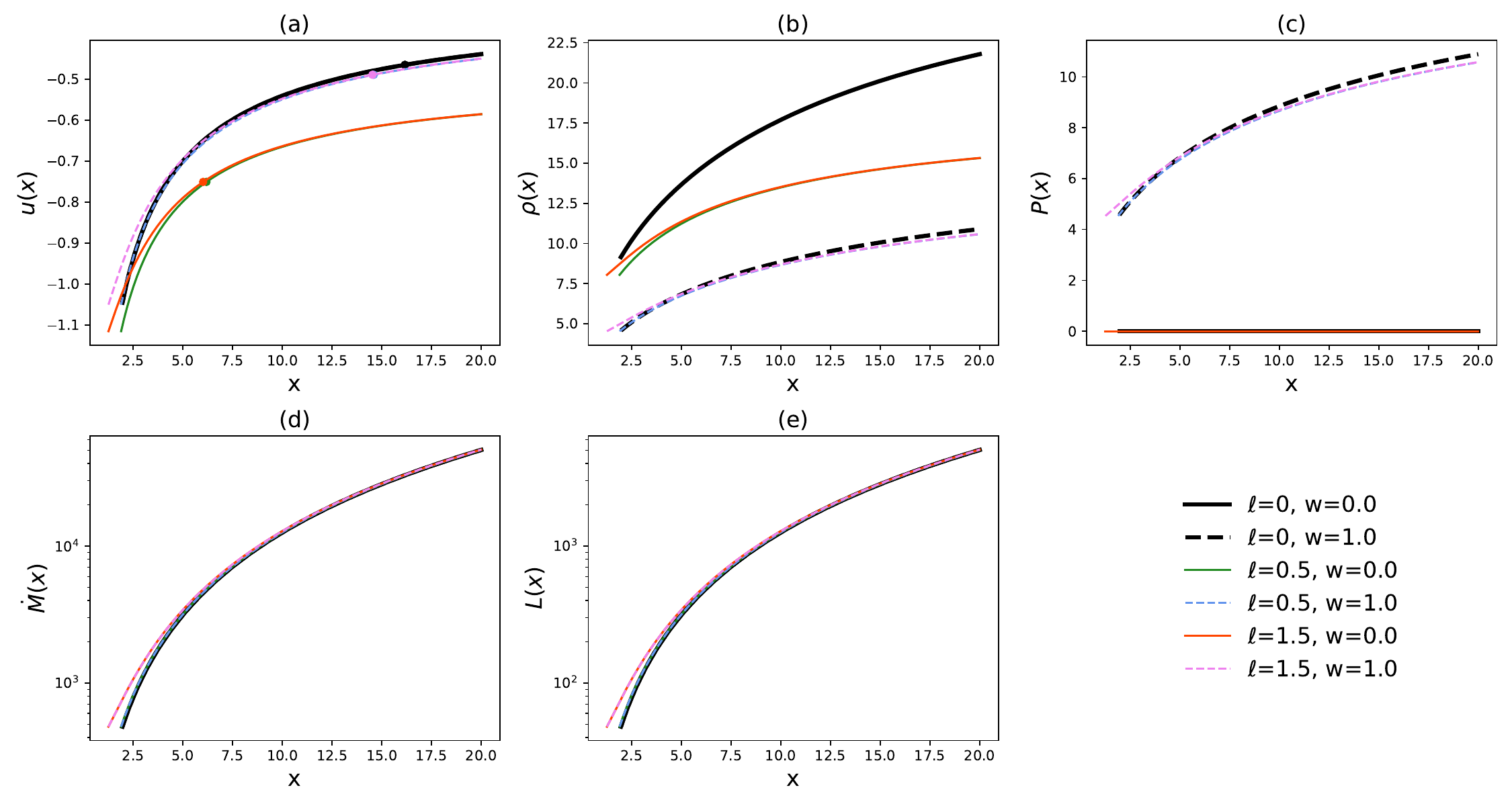}
  \caption{Comparison between RN ($\ell=0,\ Q=0.3$, black lines) and charged RBHs solutions for a barotropic EoS ($p=w\rho$) profiles \emph{w.r.t.} $x/M$: (a) radial velocity $u(x)$, (b) density $\rho(x)$, (c) pressure $P(x)$, (d) accretion rate $\dot M(x)$, and (e) luminosity $L(x)$, for $\ell=0.5,1.5$, $w=0,1$ and $Q=0.3$. Critical radii are marked with filled circles in panel (a).}
  \label{fig:RBH_RN_barotropic_comparison}
\end{figure*}

To quantify the influence of SV regularization in the charged case, we compare each configuration with the classical RN solution, namely $\ell = 0$,\ $Q = 0.3$. Figure \ref{fig:RBH_RN_barotropic_comparison} shows this comparison for both $w = 0$ and $w = 1$.

Panel (a) confirms that the inflow velocity profiles for $\ell = 0.5$ remain nearly superimposed on the RN solution throughout the domain. The SV deformation introduces negligible dynamical deviation in this regime.

Panels (b) and (c) show similar agreement in density and pressure. Differences from the RN profile are more noticeable in the stiff matter case, where the classical solution exhibits slightly steeper growth with increasing $x$. For dust, the agreement is excellent, with negligible deviations in density near the horizon. The pressure vanishes throughout for $w = 0$.

Accretion rate and luminosity, see panels (d) and (e), display very close agreement between the RN and SV solutions, with small differences emerging only near the horizon. In this region, the RN solution shows a slightly sharper drop than its regularized counterparts.
\begin{table*}[!t]
  \centering
  \renewcommand{\arraystretch}{1.2}
  \begin{tabular}{l@{\hspace{1cm}}c@{\hspace{0.5cm}}c@{\hspace{0.3cm}}c@{\hspace{0.3cm}}c}
    \hline\hline
    \multicolumn{5}{c}{\textbf{Critical point analysis: Barotropic EoS charged case}} \\
    \hline
    $(\ell,\ w)$ & $x_{\rm EH}$ & $x_c$  & $u_c$ & $\mathcal{C}_4$ \\
    \hline
    (0,\   1)  & 1.95    & 16.16   & -0.47   & 2.09  \\
    (0,\   0)  & 1.95    & 16.16   & -0.47   & 1.05  \\
    (0.5,\ 1)  & 1.89    & 14.60   & -0.49   & 2.10  \\
    (0.5,\ 0)  & 1.89    & 6.18    & -0.75   & 1.12  \\
    (1.5,\ 1)  & 1.25    & 14.53   & -0.49   & 2.10  \\
    (1.5,\ 0)  & 1.25    & 6.01    & -0.75   & 1.12  \\
    (2.5,\ 1)  & --      & 14.39   & -0.49   & 2.10  \\
    (2.5,\ 0)  & --      & 5.67    & -0.75   & 1.12  \\
    \hline\hline
  \end{tabular}
  \caption{Event horizon, sonic radius, and flow speed at the sonic point for each $(\ell,w)$ in the barotropic model with a charge of $Q=0.3$. For the analysis, we used the following values for the integration constant: $\mathcal{C}_1=10,\ \mathcal{C}_3=1.9$ and the parameters $M = 1 \text{AU},\ \eta = 0.1$.}
  \label{tab:barotropic_RN}
\end{table*}
%

\subsubsection*{Exponential density profile}
\label{RBH_RN_exponential}

\begin{figure*}[!t]
  \centering
  \includegraphics[width=0.8\textwidth]{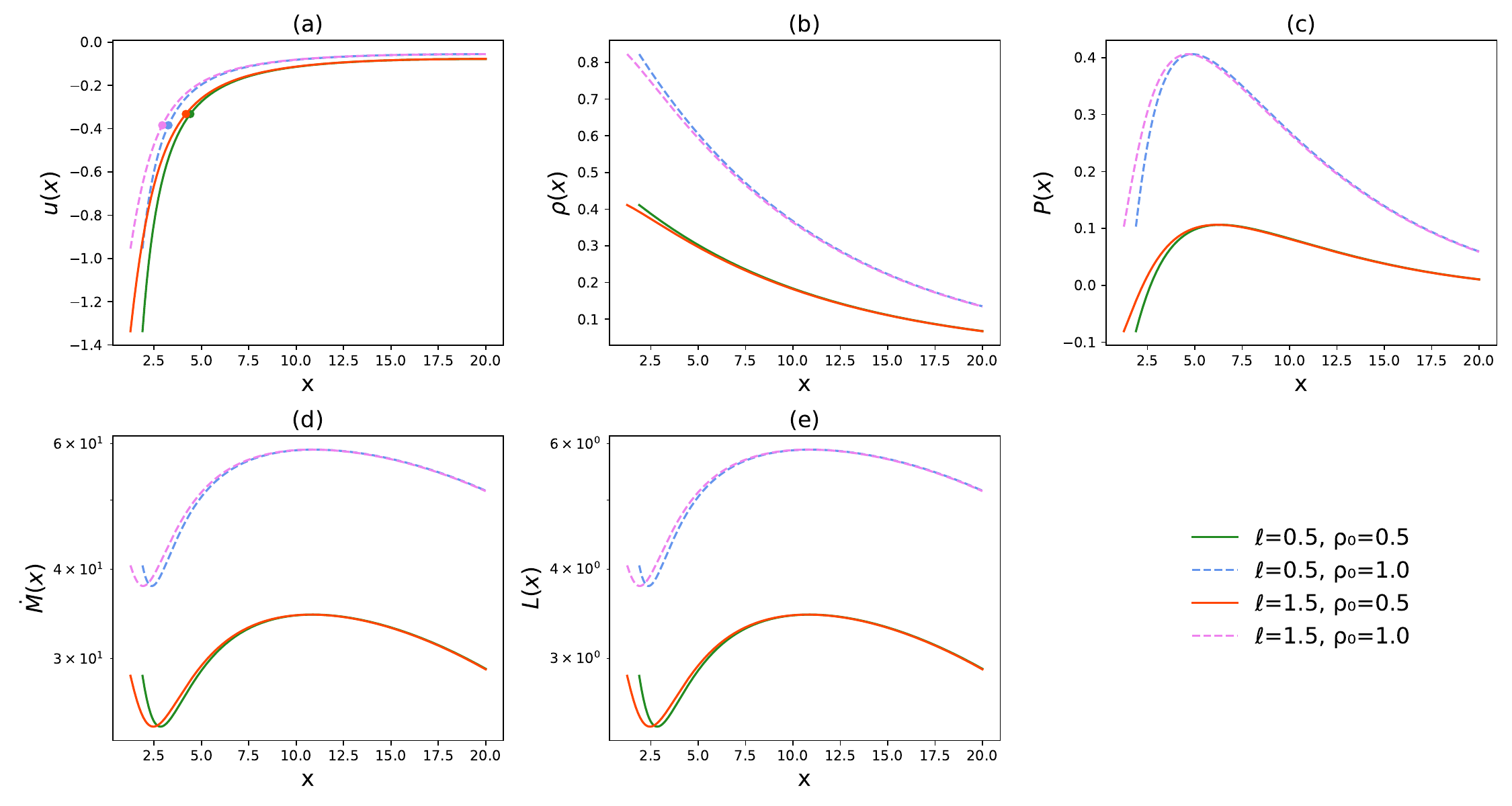}
  \caption{Exponential density profile model \emph{w.r.t.} $x/M$: (a) radial velocity $u(x)$, (b) density $\rho(x)$, (c) pressure $P(x)$, (d) accretion rate $\dot M(x)$, and (e) luminosity $L(x)$, for $\ell=0.5,1.5$,  $\rho_0=0.5,1.0\mathrm{AU}^{-2}$ and $Q=0.3$. The critical points are marked in the velocity panel.}
  \label{fig:RBH_exp_RN}
\end{figure*}

We now turn to the charged exponential density profile model. Table \ref{tab:Exp_RN} reports the critical parameters for each pair $(\ell,\rho_0)$, while Fig. \ref{fig:RBH_exp_RN} displays the radial profiles of all physical observables.

Compared with the barotropic case, the inflow velocities in panel (a) still converge at large distances but exhibit slight differences near the center, depending on $\ell$. The inclusion of charge does not significantly alter the velocity profiles, and the critical points confirm the same trend observed in the uncharged case: $x_c$ for RBHs are located closer to the center and shift outward slightly compared to the neutral configurations.

Density and pressure profiles, see panels (b) and (c), also follow similar patterns to those in the uncharged exponential density model. The characteristic split between solutions with different $\rho_0$ values is thus preserved. The density decreases exponentially with increasing $x$, while the pressure grows toward the horizon, peaks near it, and then rapidly falls off.

As expected, the accretion rate and luminosity, see panels (d) and (e), show the same structure as in the uncharged case. Starting from small $x$, both observables increase steeply near the event horizon, followed by a reduction in intermediate regions. At large $x$, the profiles rise again, reaching a maximum before decreasing asymptotically to zero as $x \rightarrow \infty$. Their major difference lies within the parameter $\eta$.

Overall, the inclusion of charge in the exponential model does not introduce substantial deviations from the uncharged case, either in profile shapes or in the values of the critical parameters. Any changes, due to the electric charge, occur only at one part over ten.

\begin{figure*}[!t]
  \centering
  \includegraphics[width=0.8\textwidth]{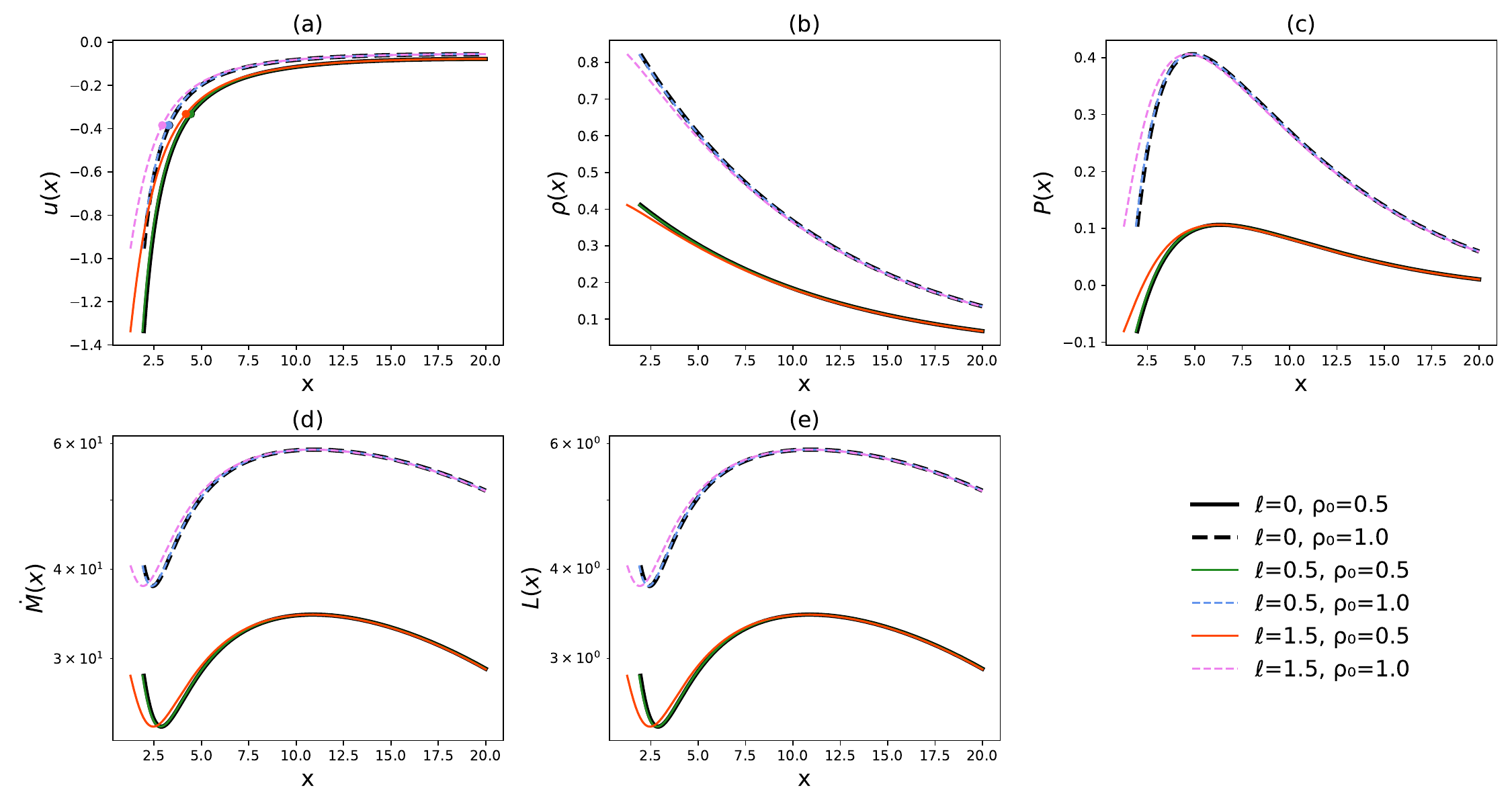}
  \caption{Comparison between RN ($\ell=0$, black lines) and RBHs solutions for an exponential density profile model \emph{w.r.t.} $x/M$: (a) radial velocity $u(x)$, (b) density $\rho(x)$, (c) pressure $P(x)$, (d) accretion rate $\dot M(x)$, and (e) luminosity $L(x)$, for ,  $\rho_0=0.5,1.0\mathrm{AU}^{-2}$ and $Q=0.3$. The critical points are marked in the velocity panel.}
  \label{fig:RBH_exp_RN_comparison}
\end{figure*}

We now compare these results with the classical RN solution with an exponential density profile, as shown in Fig. \ref{fig:RBH_exp_RN_comparison}.

Inflow velocities, see panel (a),  are nearly indistinguishable across all values of $\ell$, showing excellent agreement with the RN profile over most of the domain. A small deviation is visible for $\ell = 1.5$ near the horizon, where the RN solution falls down more steeply and closely resembles the $\ell = 0.5$ profiles.

The critical points between the RBH and RN cases are nearly coincident, though the RN values are slightly more inward compared to their uncharged Schwarzschild counterparts.

Density, pressure, mass accretion rate, and luminosity, see panels (b)--(e), display good agreement across all configurations and both $\rho_0$ values. Minor deviations arise near the event horizon, where the RN profiles show slightly stronger growth in density and pressure compared to the regularized cases. Similarly, RN displays a slightly sharper increase in both accretion rate and luminosity near the horizon. For the SV solutions, the growth is smoother and reaches,  partially deeper into the central region, due to differences in $x_{\rm EH}$.
\begin{table*}[!t]
  \centering
  \renewcommand{\arraystretch}{1.2}
  \begin{tabular}{l@{\hspace{1cm}}c@{\hspace{0.5cm}}c@{\hspace{0.3cm}}c@{\hspace{0.3cm}}c}
    \hline\hline
    \multicolumn{5}{c}{\textbf{Critical point analysis: Exponential profile - charged case}} \\
    \hline
    $(\ell,\ \rho_0)$ & $x_{\rm EH}$ & $x_c$  & $u_c$ & $\mathcal{C}_3$ \\
    \hline
    (0,\   0.5)     & 1.95      & 4.44    & -0.33  & 2.10  \\
    (0,\   1.0)     & 1.95      & 3.29    & -0.38  & 3.00  \\
    (0.5,\ 0.5)     & 1.89      & 4.41    & -0.33  & 2.10  \\
    (0.5,\ 1.0)     & 1.89      & 3.26    & -0.38  & 3.00  \\
    (1.5,\ 0.5)     & 1.25      & 4.18    & -0.33  & 2.10  \\
    (1.5,\ 1.0)     & 1.25      & 2.93    & -0.38  & 3.00  \\
    (2.5,\ 0.5)     & --        & 3.67    & -0.33  & 2.10  \\
    (2.5,\ 1.0)     & --        & 2.15    & -0.38  & 3.00  \\
    \hline\hline
  \end{tabular}
  \caption{Event horizon, sonic radius, and flow speed at the sonic point for each $(\ell,\rho_0)$ in the exponential profile model with a charge of $Q=0.3$. For the analysis, we used the following values for the integration constant: $\mathcal{C}_4=2.1$ and the parameters $M = 1 \text{AU},\ \eta = 0.1$.}
  \label{tab:Exp_RN}
\end{table*}
%

\subsection{Wormholes accretion}
\label{WH_RN_results}

We now investigate the accretion behavior in the charged wormhole regime, corresponding to the regularization parameter $\ell = 2.5$. This configuration no longer features an event horizon and instead presents a throat at $x = 0$. As in the uncharged case, we consider both barotropic and exponential fluid models, examining how the presence of electric charge modifies the accretion dynamics.

\subsubsection{Barotropic equation of state}
\label{WH_RN_barotropic}

\begin{figure*}[!t]
  \centering
  \includegraphics[width=0.8\textwidth]{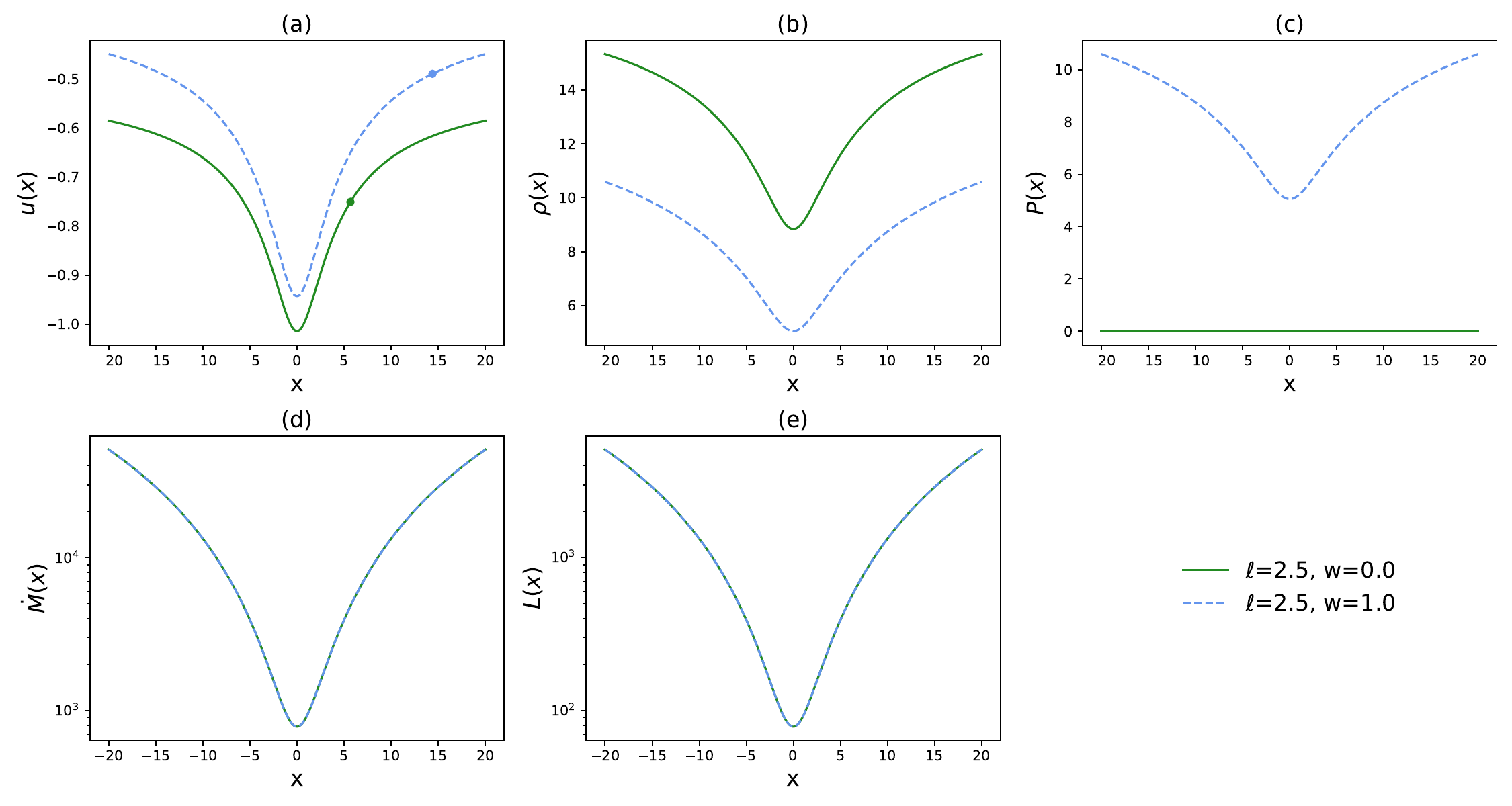}
  \caption{Barotropic EoS ($p=w\rho$) profiles \emph{w.r.t.} $x/M$: (a) radial velocity $u(x)$, (b) density $\rho(x)$, (c) pressure $P(x)$, (d) accretion rate $\dot M(x)$, and (e) luminosity $L(x)$, for $\ell=2.5$, $w=0,1$ and $Q=0.3$. Critical radii are marked with filled circles in panel (a).}
  \label{fig:WH_RN_barotropic}
\end{figure*}

Figure \ref{fig:WH_RN_barotropic} presents the charged wormhole accretion profiles for barotropic fluids.

In panel (a), the velocity profiles show the same qualitative behaviour as in the neutral wormhole case. The inflow velocities decrease symmetrically on either side of the throat and split according to the value of $w$. The critical points again differ between the stiff and dust matter cases, with the former occurring farther from the throat. Both $x_c$ values are slightly smaller compared to the neutral wormhole configuration, although still of the same order.

Density and pressure (panels (b) and (c)) remain consistent with the uncharged case. For stiff matter, the density increases with distance from the throat, while for dust, the density profile is flatter near the center. Pressure vanishes for $w = 0$, and the inclusion of charge does not alter this behaviour. For $w = 1$, the pressure follows the density profile, with slightly suppressed central values due to the presence of charge.

Accretion rate and luminosity (panels (d) and (e)) show excellent agreement with their neutral counterparts. All solutions overlap at large distances, with only minor deviations near the throat. In particular, the RN solution displays a more abrupt decline near the center, while the wormhole profiles remain smooth and symmetric.

\begin{figure*}[!t]
  \centering
  \includegraphics[width=0.8\textwidth]{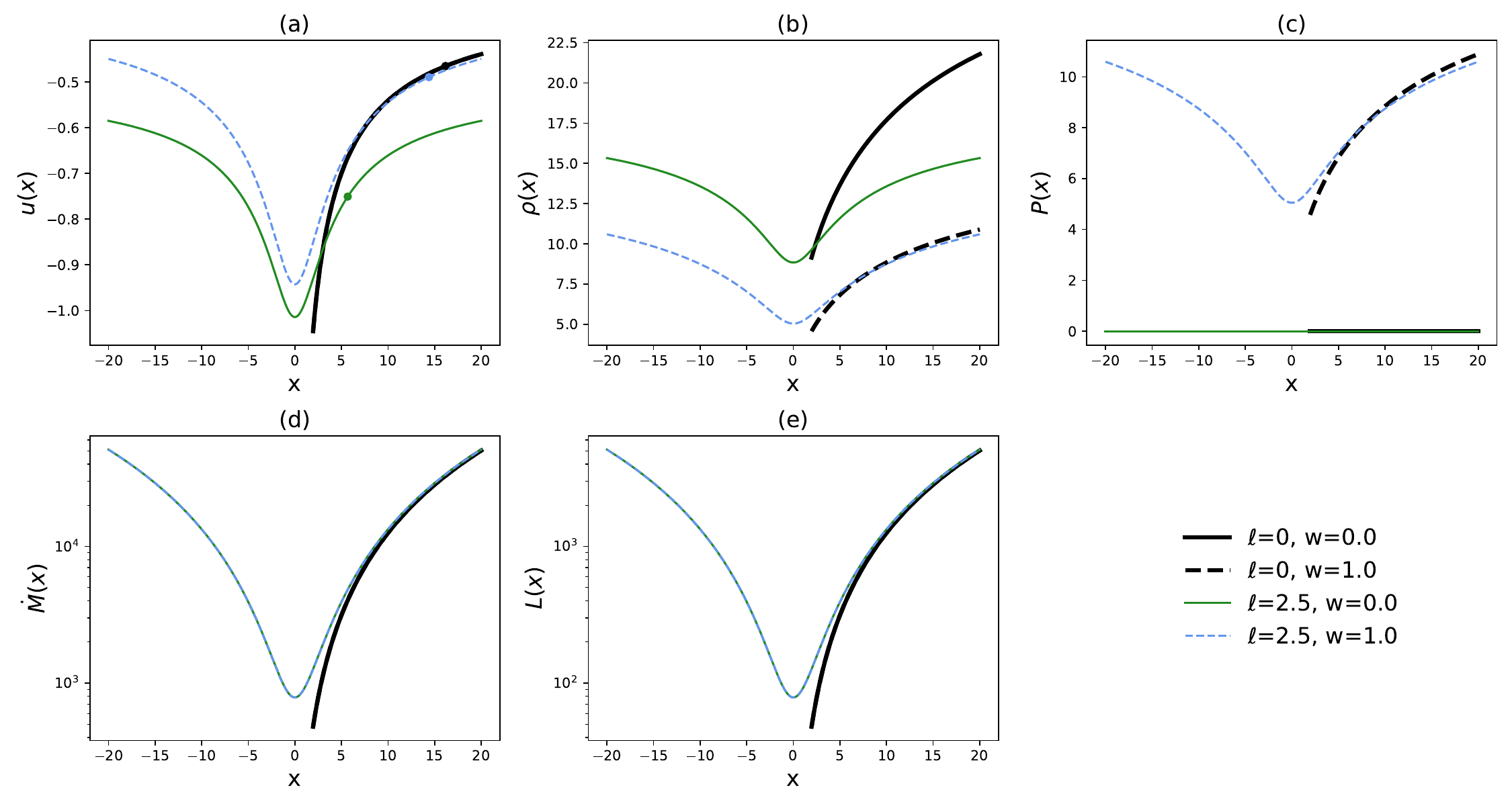}
  \caption{Comparison between RN ($\ell=0,\ Q=0.3$, black lines) and charged wormholes solutions for a barotropic EoS ($p=w\rho$) profiles \emph{w.r.t.} $x/M$: (a) radial velocity $u(x)$, (b) density $\rho(x)$, (c) pressure $P(x)$, (d) accretion rate $\dot M(x)$, and (e) luminosity $L(x)$, for $\ell=2.5$, $w=0,1$ and $Q=0.3$. Critical radii are marked with filled circles in panel (a).}
  \label{fig:WH_RN_barotropic_comparison}
\end{figure*}

Figure \ref{fig:WH_RN_barotropic_comparison} compares these results with the classical RN geometry.

Inflow velocities for stiff matter nearly coincide with the RN solution, including at the critical point. For dust matter, deviations are visible near the throat, where the wormhole maintains a slower, more gradual profile. This pattern persists in the density panel: RN grows more steeply than the wormhole. The pressure instead, shows good agreement throguh the entire range of $x$.

The accretion rate and luminosity for all solutions agree in the outer region, diverging only near the throat where RN shows a sharper dip compared to the regularised models.

\subsubsection{Exponential density profile}
\label{WH_RN_exponential}

\begin{figure*}[!t]
  \centering
  \includegraphics[width=0.8\textwidth]{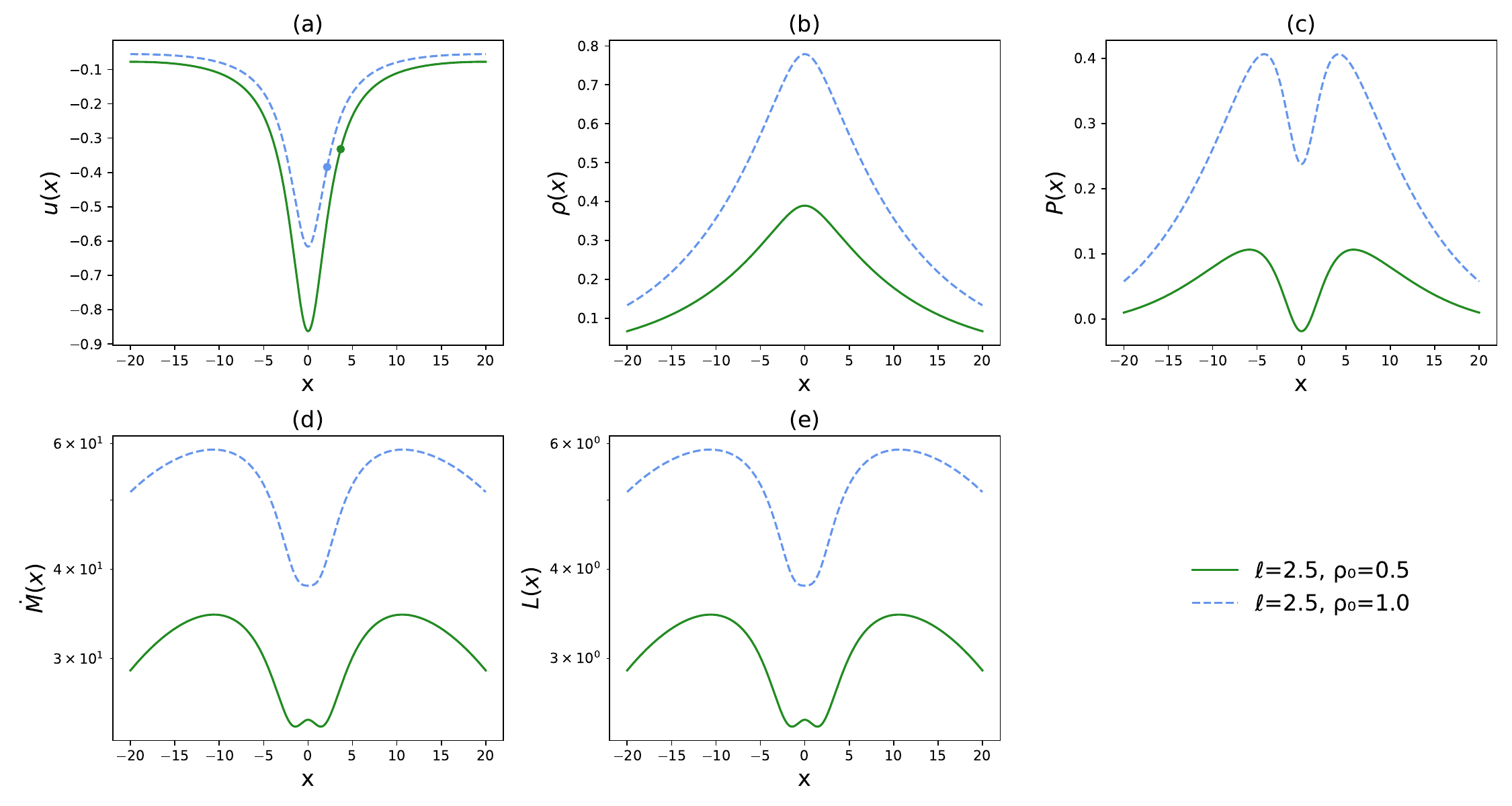}
  \caption{Exponential density profile model \emph{w.r.t.} $x/M$: (a) radial velocity $u(x)$, (b) density $\rho(x)$, (c) pressure $P(x)$, (d) accretion rate $\dot M(x)$, and (e) luminosity $L(x)$, for $\ell=2.5$,  $\rho_0=0.5,1.0\mathrm{AU}^{-2}$ and $Q=0.3$. The critical points are marked in the velocity panel.}
  \label{fig:WH_exp_RN}
\end{figure*}

Figure \ref{fig:WH_exp_RN} shows the exponential accretion solutions for the charged wormhole configuration.

The velocity profiles in panel (a) remain largely unaffected by the presence of charge, with trends mirroring the neutral case. Due to the absence of a horizon, velocities are smooth across the throat. However, the critical points in this configuration exhibit more significant differences, with values noticeably smaller compared to both neutral and charged RBH solutions. This highlights the more substantial geometric alteration introduced by the wormhole topology.

Density and pressure profiles (panels (b) and (c)) behave as expected. For both values of $\rho_0$, the density decreases with $x$, while the pressure increases towards the throat before sharply decreasing. The maximum values reached are higher for $\rho_0 = 1$, consistent with the uncharged case. The electric charge introduces only mild modifications to these profiles, largely confined to the innermost region.

Accretion rate and luminosity (panels (d) and (e)) are qualitative not altered by the presence of $Q$. Starting from small $x$, both observables grow toward the throat, then drop as $x$ increases, and later rise again before fading out asymptotically. As in the uncharged scenario, the inclusion of charge produces small numerical shifts but does not alter the overall behaviour.

\begin{figure*}[!t]
  \centering
  \includegraphics[width=0.8\textwidth]{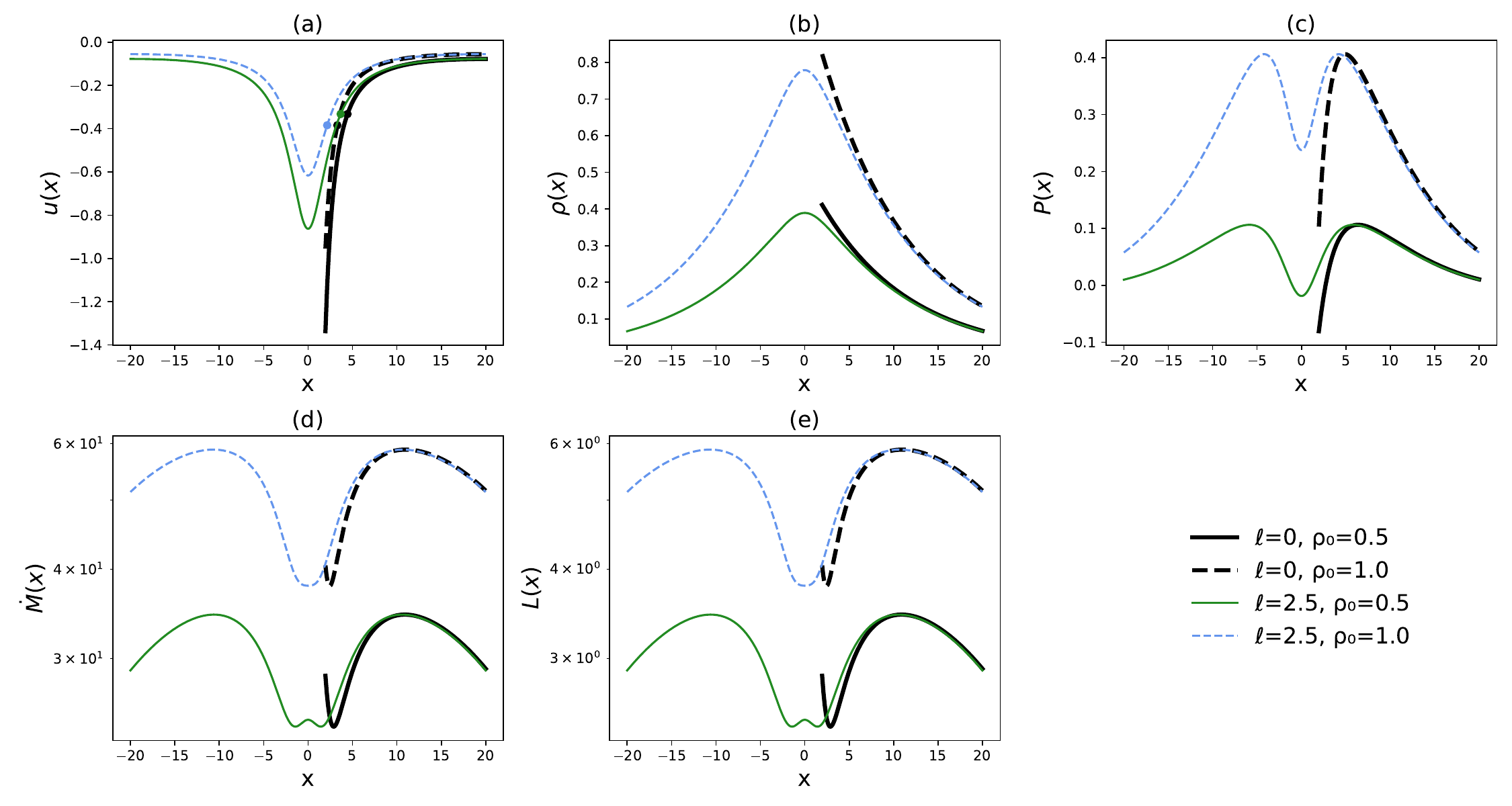}
  \caption{Comparison between Schwarzschild ($\ell=0,\ Q=0.3$, black lines) and charged wormholes solutions for an exponential density profile model \emph{w.r.t.} $x/M$: (a) radial velocity $u(x)$, (b) density $\rho(x)$, (c) pressure $P(x)$, (d) accretion rate $\dot M(x)$, and (e) luminosity $L(x)$, for $\ell=2.5$, $\rho_0=0.5,1.0\mathrm{AU}^{-2}$ and $Q=0.3$. The critical points are marked in the velocity panel.}
  \label{fig:WH_exp_RN_comparison}
\end{figure*}

Figure \ref{fig:WH_exp_RN_comparison} compares these results with the RN baseline. While the profiles largely overlap at large $x$, near the throat the RN solution exhibits more abrupt gradients. The RN velocity decreases more sharply than its wormhole counterparts. The critical point lies farther out in the RN solution than in the charged wormhole case.

For the other variables, all configurations agree well at large radii, but the RN solution deviates in the inner region. Density and pressure grow more rapidly in RN, and similarly, the RN accretion rate and luminosity show stronger peaks near the center. In contrast, the wormhole solutions display smoother transitions with slower growth.

\subsubsection{Deviations from the Reissner-Nordstr\"om solution}
\label{result_dev_RB}

For the barotropic model, the critical radius $x_c$ shows deviations below $1\%$ for RBHs and $\sim1\%$ for the wormhole regimes with $w=0$ and $\sim 8\%$ for the other case. The location of the event horizon shifts by up to $36\%$ for $\ell = 1.5$. In both dust and stiff matter cases, the critical velocity $u_c$ remains constant to numerical precision. These results suggest that the dynamics of the accreting fluid are only weakly sensitive to the SV regularization.

In the exponential model, deviations are more pronounced. For example, $x_c$ shifts by $\sim 5-10\%$ for $\ell = 1.5$ and reaches $\sim 35\%$ for $\ell = 2.5$. Yet, the critical velocity remains unchanged across all configurations. Thus, while the presence of charge and the SV deformation significantly impact the spatial location of the critical point, they do not substantially alter the dynamical inflow behaviour. The most significant effects are observed in the wormhole regime near the throat, where the structure of the spacetime diverges more strongly from the classical RN geometry.

\begin{table*}[!t]
  \centering
  \renewcommand{\arraystretch}{1.2}
  \begin{tabular}{l c c c c c}
    \hline\hline
    Model & Param & $\ell$ & $\Delta x_{EH}[\%]$ & $\Delta x_c[\%]$ & $\Delta u_c[\%]$ \\
    \hline\hline
    \multirow{3}{*}{Barotropic EoS, $w=0$}
      &                 & 0.5   & $3.33$        & $0.33$  & $0$     \\
      &                 & 1.5   & $35.92$       & $2.97$  & $0$     \\
      &                 & 2.5   & ---           & $8.49$  & $0$     \\
    \hline
    \multirow{3}{*}{Barotropic EoS, $w=1$}
      &                 & 0.5   & $3.33$        & $0.06$  & $0$     \\
      &                 & 1.5   & $35.92$       & $0.53$  & $0$     \\
      &                 & 2.5   & ---           & $1.48$  & $0$     \\
    \hline
    \multirow{3}{*}{Exponential, $\rho_0=0.5$}
      &                 & 0.5   & $3.33$        & $0.64$  & $0$     \\
      &                 & 1.5   & $35.92$       & $5.88$  & $0$     \\
      &                 & 2.5   & ---           & $17.36$ & $0$     \\
    \hline
    \multirow{3}{*}{Exponential, $\rho_0=1.0$}
      &                 & 0.5   & $3.33$        & $1.16$  & $0$     \\
      &                 & 1.5   & $35.92$       & $10.96$ & $0$     \\
      &                 & 2.5   & ---           & $34.86$ & $0$     \\
    \hline\hline
  \end{tabular}
  \caption{Percentage deviations of the horizon coordinate $x_{EH}$, sonic radius $x_c$, and critical velocity $u_c$ for $\ell>0$ in the charged SV ($Q=0.3$) spacetime, relative to the Reissner-Nordstr\"om solution. Entries “$---$” indicate no horizon in the wormhole geometry ($\ell=2.5$).}
  \label{tab:percent_dev_RN}
\end{table*}

%
%
\section{Final outlooks}
\label{CONCLUSION}

In this work, we performed a detailed analysis of Bondi-type spherical accretion within the frameworks of SV spacetimes, including different classes of solutions.

In particular, our first approach considers a neutral family of spacetimes, whereas the second consists of adding an electric charge $Q$, appearing as physical electromagnetic charge, different from topological contribution, as, for example, provided by the Bardeen solution.

We assessed whether different matter models can induce observationally distinguishable features when accreting onto different geometries, as opposed to standard BHs. Hence, we worked out two distinct fluid configurations.

In the first case, we focused on a barotropic fluid characterized by the EoS $P = w\,\rho$, with $w$ constant, representing a broad class of accreting matter types, inspired by cosmology. Particularly, the extreme cases of dust, $w=0$, and stiff matter, $w=1$, have been explored. In the second class of fluids, we focused on an exponential radial density profile, previously employed in the context of dark matter accretion models \cite{Sofue:2013a,Sofue:2013b,Boshkayev:2020}, dubbed \emph{Sofue or exponential profile}.

Both fluid models have been applied to a representative set of solutions, as stated above and, in particular, we exploited RBHs and wormhole solutions embedded in the SV framework, allowing us to compute key accretion observables such as the mass accretion rate, the location of the critical (or sonic) radius, and the emergent luminosity.

Starting from the neutrally charged SV solutions, our analysis showed that, for RBHs solutions ($0  <\ell\leq 2$), while the Bondi formalism remains structurally stable under the exponential density profile, the barotropic case exhibits non-trivial modifications in the position of the critical point, as well as in the qualitative behaviour of both the mass flux and the associated luminosity. This is also perpetuated in wormhole solutions ($\ell>2$) with the only difference that the trends are symmetrically the same also for negative $x$ since this solution does not have an event horizon but only a throat at $x=0$.

Afterwards, by comparing these configurations with their classical counterparts, we saw that for RBH the biggest deviations appeared more evident in the inner part of the accretion disk, near the event horizon, and mildly in the results of the critical point analysis. The classical solution is shown to become supersonic before the SV solutions, a fact that applies to all $\ell$, in both cases of exponential density profile and barotropic fluid. We pursued these findings with the aim of characterizing potential deviations from the Schwarzschild solution. In particular, these deviations have been found to be remarkable in precise regions, suggesting that possible future measurements can distinguish RBH solutions from the Schwarzschild case by measuring the free parameters of the underlying metrics.

Immediately after the neutral scenarios, we then extended our work including $Q$, implying a class of charged SV solutions. In this case, we used a moderate electric charge value, namely $Q = 0.3$, and compared the resulting accretion profiles with those predicted by the RN geometry. The impact of $Q$ was found to be mild for both values of $\ell$ and fluid models.
Starting from the RBHs solutions, the inclusion of charge led to minor shifts in the critical radii and small quantitative changes in the fluid observables. These deviations were generally limited to decimals and became more evident only in the inner regions. More precisely, the largest differences were observed in the density profiles of the stiff matter case and in the position of the sonic point for exponential fluids with high central densities. For what concerns the wormholes solutions instead, we have the biggest deviations from the uncharged case showing that, as the value of $\ell$ increases, the differences from the classical solution become more pronounced especially in the positions of the critical points.

Nevertheless, for both RBH and wormhole configurations, the presence of charge produced limited effects. Both the barotropic and exponential models were in agreement with the trend of their uncharged counterparts. Critical radii experience slightly shifts, but the fluid dynamics remained regular. When compared with the classical RN solution, the solutions deviate noticeably only near the throat or the horizon, except the density profile for the stiff matter case.

Summing up, the regularization parameter $\ell$ remains the dominant driver of geometric deviations, while the inclusion of moderate electric charge acts as a small perturbation of  critical quantities. This fact indicates that a further \emph{hairy} parameter, in this case the charge $Q$, within SV solutions, does not particularly influence the analysis.

This opens to possible future investigations of additional parameters that, instead, may influence, more significantly, the overall expectations.

In this respect, we therefore intend to explore more complicated cases of SV solutions, involving complicated, and alternative, versions of the SV spacetime. Further, the role played by this class of solutions will be  explored more deeply concerning its connection to particle-configurations, see e.g. \cite{Luongo:2025iqq}, and therefore to describe dark matter. Our future studies will also shed light on possible new observable quantities, as well as SV thermodynamics.

\section*{Acknowledgments}
SG and RG are thankful to Salvatore Capozziello for insightful discussions related to the physics of accretion disks in theories that depart from general relativity. OL thanks Abraao Jesse Capistrano de Souza and Daniele Malafarina for interesting debates on the physics of accretion disk and acknowledges financial support from the Fondazione  ICSC, Spoke 3 Astrophysics and Cosmos Observations. National Recovery and Resilience Plan (Piano Nazionale di Ripresa e Resilienza, PNRR) Project ID CN$\_$00000013 "Italian Research Center on  High-Performance Computing, Big Data and Quantum Computing"  funded by MUR Missione 4 Componente 2 Investimento 1.4: Potenziamento strutture di ricerca e creazione di "campioni nazionali di R$\&$S (M4C2-19 )" - Next Generation EU (NGEU)
GRAB-IT Project, PNRR Cascade Funding
Call, Spoke 3, INAF Italian National Institute for Astrophysics, Project code CN00000013, Project Code (CUP): C53C22000350006, cost center STI442016. The authors are also grateful to Mohammad Bagher Jahani Poshteh and Marco Muccino for their valuable help in computational stuff.

\clearpage
\appendix
\onecolumngrid
\section{Numerical trends for charged and uncharged Simpson-Visser solutions}
\label{appendix}
Here, we have collected all the figures showing the different trends of the two families of solutions, i.e. charged and uncharged Simpson-Visser spacetimes, for the various compact objects analysed throughout this work.
%
%
%
%
%
%
%
%
%

\end{document}